\documentclass[12pt,a4paper]{article}
\usepackage[british]{babel}
\usepackage{epsfig}
\begin{document}
\date{}

\title{
{\vspace{-20mm} \normalsize
\hfill \parbox[t]{50mm}{DESY 04-162     \\
                        MS-TP-04-27     \\
                        SFB/CPP-04-56}} \\[20mm]
 The phase structure of lattice QCD with \\
 two flavours of Wilson quarks and \\
 renormalization group improved gluons \\[5mm]}

\author{F.\ Farchioni$^{a}$,
        K.\ Jansen$^{b}$,
        I.\ Montvay$^{c}$,
        E.\ Scholz$^{c}$,
        L.\ Scorzato$^{d}$,\\
        A.\ Shindler$^{b}$,
        N.\ Ukita$^{c}$,
        C.\ Urbach$^{b,e}$,
        I.\ Wetzorke$^{b}$\\[5mm]
  {\small $^a$ Universit\"at M\"unster,
   Institut f\"ur Theoretische Physik,}\\
   {\small Wilhelm-Klemm-Strasse 9, D-48149 M\"unster,
   Germany}\\
  {\small $^b$ NIC/DESY Zeuthen, Platanenallee 6, D-15738 Zeuthen,
   Germany}\\
  {\small $^c$ Deutsches Elektronen-Synchrotron DESY, Notkestr.\,85,
   D-22603 Hamburg, Germany}\\
  {\small $^d$ Institut f\"ur Physik, Humboldt Universit\"at zu Berlin,
   D-12489 Berlin, Germany}\\
  {\small $^e$ Freie Universit\"at Berlin,
   Institut f\"ur Theoretische Physik,}\\
  {\small Arnimallee 14, D-14196 Berlin, Germany}}
%
\newcommand{\be}{\begin{equation}}                                              
\newcommand{\ee}{\end{equation}}                                                
\newcommand{\half}{\frac{1}{2}}                                                 
\newcommand{\rar}{\rightarrow}                                                  
\newcommand{\lar}{\leftarrow}
\newcommand{\LCB}{\raisebox{-0.3ex}{\mbox{\LARGE$\left\{\right.$}}}
\newcommand{\RCB}{\raisebox{-0.3ex}{\mbox{\LARGE$\left.\right\}$}}}
\newcommand{\U}{\mathrm{U}}
\newcommand{\SU}{\mathrm{SU}}
\newcommand{\bteq}[1]{\boldmath$#1$\unboldmath}

 
\maketitle

\abstract{
 The effect of changing the lattice action for the gluon field
 on the recently observed \protect\cite{TWIST} first order phase
 transition near zero quark mass is investigated by replacing the Wilson
 plaquette action by the DBW2 action.
 The lattice action for quarks is unchanged: it is in both cases the
 original Wilson action.
 It turns out that Wilson fermions with the DBW2 gauge action have
 a phase structure where the minimal pion mass and the jump
 of the average plaquette are decreased, when compared to Wilson fermions
 with Wilson plaquette action at similar values of the lattice spacing.
 Taking the DBW2 gauge action is advantageous also from the point of
 view of the computational costs of numerical simulations.}

\newpage
\section{Introduction}\label{sec1}

 A basic feature of the low energy dynamics in Quantum Chromodynamics
 (QCD) is the spontaneous chiral symmetry breaking implying the
 existence of light pseudo-Goldstone (pseudoscalar) bosons.
 The associated phase structure near zero quark masses has to be
 reproduced in the continuum limit by the lattice-regularized
 formulations but it is in general modified by lattice artifacts at
 non-vanishing lattice spacing.
 In lattice theories based on Wilson-type quark actions the possible
 phase structures have been investigated up to ${\cal O}(a^2)$ in the
 lattice spacing $a$ by Sharpe and Singleton \cite{SHARPE-SINGLETON} in
 the framework of low-energy chiral Lagrangians \cite{WEINBERG,CHPT} and
 using the effective continuum description of cut-off effects
 \cite{SYMANZIK,CLOVER}.
 Their results allow two possible ``scenarios'': the existence of
 the Aoki phase \cite{AOKI-PHASE} or, alternatively, a first order
 phase transition near zero quark mass.

\begin{figure}
%
\begin{center}
{\Large\bf Figures}
\end{center}
\begin{center}
\vspace*{0.10\vsize}
\begin{minipage}[c]{1.25\linewidth}
\includegraphics[angle=-90,width=.95\hsize]
 {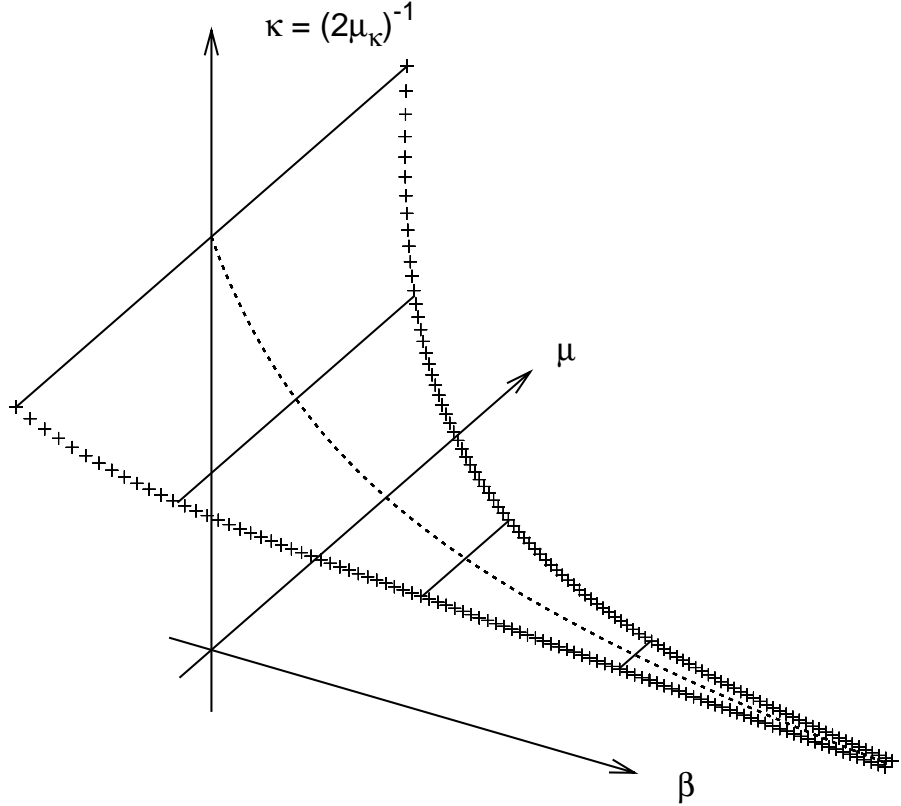}
\end{minipage}
\end{center}
\vspace*{-0.15\vsize}
\begin{center}
\parbox{0.8\linewidth}{\caption{\label{fig_phasediag}\em
 The schematic view of the first order phase transition surface
 in the $(\beta,\kappa,\mu)$ space close to the continuum limit.
 ($\beta$=bare gauge coupling, $\kappa$=hopping parameter,
 $\mu$=bare twisted quark mass, $\mu_\kappa \equiv (2\kappa)^{-1}$=bare
 untwisted quark mass.)
 The crosses mark the second order boundary line of the first order
 phase transition surface.
 The strong coupling region near $\beta=0$ is not shown in this
 figure.}}
\end{center}
\end{figure}

 In a recent numerical simulation \cite{TWIST,FERMILPROC} the phase
 structure of lattice QCD with Wilson fermions and Wilson gauge action
 has been investigated with the help of the twisted mass Wilson fermion
 formulation \cite{TMQCD,ALPHA-TMQCD}.
 For fixed values of $a$, smaller than $a \approx 0.2\,{\rm fm}$,
 evidence for a first order phase transition line, near zero quark mass
 in the plane of untwisted and twisted quark mass, has been found
 corresponding to the ``second scenario'' of
 ref.~\cite{SHARPE-SINGLETON}.
 It is important to remark that this line is finite and ends at a
 particular value of the twisted quark mass $\mu_c$.
 This implies metastability and a non-zero minimum of the absolute value
 of quark- (and pion-) masses.
 These are lattice artifacts which are expected to vanish in the
 continuum limit where $\mu_c=0$ and the first order phase transition
 line shrinks to a singular point.
 (For generalizations of the results of \cite{SHARPE-SINGLETON} for
 non-zero twisted mass see \cite{MUENSTER,SHARPE-WU,SCORZATO}.)
 Considering, besides the bare quark masses, the bare gauge coupling,
 too, near the continuum limit the first order phase transition spans a
 surface, as it is schematically shown by figure~\ref{fig_phasediag}.

 It might be speculated that at the microscopic level the occurence of
 the first order phase transition at $a >0$ is accompanied by a
 massive rearrangement of small eigenvalues of the Wilson-Dirac
 operator.
 The detailed properties and, in particular, the strength of the first
 order phase transition does probably depend on the number and
 distribution of these eigenvalues.
 It is known that some type of small eigenvalues, especially real ones,
 are associated with small topological dislocations of the gauge
 field.
 A high probability of these dislocations and of the correspoding small
 eigenvalues is presumably a cut-off effect which can be diminished by
 an appropriate choice of the lattice action.
 In fact, it is known
 \cite{DOMAINWALL,RBC-PROC,RBC-PHYSREV,JANSEN-NAGAI} that the small
 topological dislocations can, indeed, be suppressed by taking
 renormalization group improved (RGI) gauge actions as the
 Iwasaki-action \cite{IWASAKI} or the DBW2 action \cite{DBW2}.

 In the present paper we try to answer the question whether the
 combination of RGI gauge actions with the Wilson fermion action
 does shrink the first order phase transition line near zero quark mass.
 Here we restrict ourselves to the study of the DBW2 gauge action
 which has been successfully applied also in dynamical domain wall
 fermion simulations \cite{DYNAMICALDW}.
 The goal of the present paper is to qualitatively show how a change of
 the gauge action will modify the phase structure.
 Hence, we do not aim here at a high precision study.

 The Iwasaki action is often used in dynamical quark simulations by
 the CP-PACS and JLQCD Collaborations, in particular, in combination
 with the Sheikholeslami-Wohlert clover-improved Wilson fermion
 action \cite{CLOVER}.
 Earlier results of the JLQCD Collaborations indicate \cite{JLQCD}
 that, indeed, a metastability seen in the average plaquette can be
 suppressed by replacing the Wilson plaquette action by the
 Iwasaki action.
 (See also \cite{JLQCD-NEW}, and for a review of earlier results on
 the phase structure of QCD \cite{UKAWA}.
 An early discussion of the phase structure of QCD can also be found
 in \cite{CREUTZ}.)

 The plan of this paper is as follows: in the next section the lattice
 action and some parameters of the update algorithm are defined.
 In section \ref{sec3} we present the results of the numerical
 simulations.
 Section \ref{sec4} is devoted to the investigation of the eigenvalue
 spectrum of the Wilson-Dirac operator near the origin.
 The last section contains some discussion and concluding remarks.

\section{Lattice action and simulation algorithm}\label{sec2}
\subsection{Lattice action}\label{sec2.1}

 We apply for quarks the lattice action of Wilson fermions, which can be
 written as
\be\label{eq2:01}
S_q = \sum_x \left\{ 
\left( \overline{\chi}_x [\mu_\kappa + i\gamma_5\tau_3\mu ]\chi_x \right)
- \half\sum_{\mu=\pm 1}^{\pm 4}
\left( \overline{\chi}_{x+\hat{\mu}}U_{x\mu}[r+\gamma_\mu]\chi_x \right)
\right\} \ .
\ee
 Here the (``untwisted'') bare quark mass in lattice units is
 denoted by
\be\label{eq2:02}
\mu_\kappa \equiv am_0 + 4r = \frac{1}{2\kappa} \ ,
\ee
 $r$ is the Wilson-parameter, set in our simulations to $r=1$, $am_0$
 is another convention for the bare quark mass in lattice units and
 $\kappa$ is the conventional hopping parameter.
 In (\ref{eq2:01}) the twisted mass $\mu$ is also introduced.
 $U_{x\mu} \in {\rm SU(3)}$ is the gauge link variable and we also
 defined $U_{x,-\mu} = U_{x-\hat{\mu},\mu}^\dagger$ and
 $\gamma_{-\mu}=-\gamma_\mu$.

 For the SU(3) Yang-Mills gauge field we apply the DBW2 lattice action
 \cite{DBW2} which belongs to a one-parameter family of actions
 obtained by renormalization group considerations.
 These actions also include, besides the usual $(1\times 1)$ Wilson loop
 plaquette term, planar rectangular $(1\times 2)$ Wilson loops:
\be\label{eq2:03}
S_g = \beta\sum_{x}\left(c_{0}\sum_{\mu<\nu;\,\mu,\nu=1}^4
\left\{1-\frac{1}{3}\,{\rm Re\,} U_{x\mu\nu}^{1\times 1}\right\}
+c_{1}\sum_{\mu\ne\nu;\,\mu,\nu=1}^4
\left\{1-\frac{1}{3}\,{\rm Re\,} U_{x\mu\nu}^{1\times 2}\right\}
\right) \ ,
\ee
 with the normalization condition $c_{0}=1-8c_{1}$.
 (The notation $c_{0,1}$ is conventional.
 Of course, $c_1$ should not be confused with the parameter $c_1$
 in the effective potential of refs.~\cite{SHARPE-SINGLETON,TWIST}.)
 The coefficient $c_{1}$ in (\ref{eq2:03}) takes different values for
 various choices of RGI actions, for instance,
\be\label{eq2:04}
c_{1} =\left\{\begin{array}{ll}
-0.331 & \textrm{Iwasaki action,} \\[0.3em]
-1.4088 & \textrm{DBW2 action.}
\end{array}\right.
\ee
 Clearly, $c_1=0$ corresponds to the original Wilson gauge action with
 the plaquette term only.
 Note that for $c_1 = -1/12$ one obtains the tree-level improved action
 in the Symanzik-improvement scheme \cite{WEISZ-WOHLERT}.

\subsection{Twist angle}\label{sec2.2}

 An important quantity is the twist angle $\omega$, the polar angle in
 the plane of the untwisted and twisted mass $(\mu_\kappa,\mu)$.
 We present here a method which allows to determine the twist angle only
 on the basis of symmetry of the correlators defined in a given point of
 bare parameter space (see also \cite{FERMILPROC}).

 Following~\cite{ALPHA-TMQCD}, we introduce the twist angle $\omega$ as
 the chiral rotation angle between the renormalized (physical) vector
 and axialvector currents $\hat{V}^a_{x\mu}$, $\hat{A}^a_{x\mu}$  and
 the bare bilinears of the $\chi$-fields $V^a_{x\mu}$, $A^a_{x\mu}$:
\be\label{eq2:05}
V^a_{x\mu} \equiv \overline{\chi}_x \half\tau_a\gamma_\mu \chi_x \ ,
\hspace{3em}
A^a_{x\mu} \equiv \overline{\chi}_x \half\tau_a\gamma_\mu\gamma_5 
\chi_x \ .
\ee
 With the renormalization constants $Z_V$ and $Z_A$ we have
\begin{eqnarray}\label{eq2:06}
& & \hat{V}^a_{x\mu} = Z_V V^a_{x\mu}\,\cos\omega\, + 
\epsilon_{ab} \, Z_A A^b_{x\mu}\,\sin\omega \ ,
\\[0.5em]\label{eq2:07}
& & \hat{A}^a_{x\mu} = Z_A A^a_{x\mu}\,\cos\omega\, +
\epsilon_{ab} \, Z_V V^b_{x\mu}\,\sin\omega\,
\end{eqnarray}
 where only charged currents are considered ($a$=1,2). 

 The twist angle $\omega$ is related to the ratio of the renormalized
 twisted and untwisted masses entering the chiral Ward
 identities~\cite{ALPHA-TMQCD}.
 (In~\cite{ALPHA-TMQCD} this definition of the twist angle was called
 $\alpha$.)
 We define, in addition, the two auxiliary angles
\be\label{eq2:08}
\omega_V=\arctan (Z_AZ_V^{-1}\tan\omega)\ ,
\hspace{3em}
\omega_A=\arctan (Z_VZ_A^{-1}\tan\omega)\ .  
\ee
 In terms of $\omega_V$, $\omega_A$ eqs.~(\ref{eq2:06}) and
 (\ref{eq2:07}) are written as
\begin{eqnarray}\label{eq2:09}
& & \hat{V}^a_{x\mu} = \!\ {\cal N}_V\, (\cos\omega_V V^a_{x\mu} +
\epsilon_{ab}\sin{\omega}_{V} A^b_{x\mu})\ ,
\\\label{eq2:10}
& & \hat{A}^a_{x\mu}=\!\ {\cal N}_A\, (\cos\omega_A A^a_{x\mu} +
\epsilon_{ab}\sin{\omega}_{A} V^b_{x\mu})
\end{eqnarray}
 where the overall multiplicative renormalization is ($X=V,A$):
\be\label{eq2:11}
{\cal N}_{X}=\frac{Z_X}
{\cos\omega_X \sqrt{1+\tan{\omega}_V\tan{\omega}_A}}\ .
\ee
 From (\ref{eq2:08}) it follows that
\be\label{eq2:12}
\omega=\arctan \left(\sqrt{\tan{\omega}_V\tan{\omega}_A}\right)\ .
\ee
 As shown by the relations in (\ref{eq2:08}) and (\ref{eq2:12}),
 the values of $\omega$, $\omega_V$ and $\omega_A$ coincide for
 $|\omega| = 0,\pi/2$.
 However, for other angles they are, in general, different and the
 difference goes to zero in the continuum limit only as fast as
 $Z_V/Z_A \to 1$.

 A possibility to determine $\omega_V$ and $\omega_A$ is to impose
 the vector and axialvector Ward identities, respectively,
 with a suitable insertion operator $\hat{O}_x$.
 For instance, in the vector case one can use the Ward identity
\be\label{eq2:13}
\sum_{\vec{x},\vec{y}} 
\langle \partial^\ast_\mu \hat{V}^+_{x\mu}\, \hat{O}^-_y\rangle = 0
\hspace*{3em}\Longrightarrow\hspace*{3em}
\tan{\omega}_V =
\frac{-i\sum_{\vec{x},\vec{y}}\langle
      \partial^\ast_0 V^+_{x0}\, \hat{O}^-_y \rangle}
     {  \sum_{\vec{x},\vec{y}}\langle
      \partial^\ast_0 A^+_{x0}\, \hat{O}^-_y \rangle} \ .
\ee
 Here the indices $+$ and $-$ refer to the charged components
 $\tau_\pm \equiv \tau_1 \pm i\tau_2$ and $\partial^\ast_\mu$
 denotes the backward lattice derivative.

 Another possibility for determining the twist angles $\omega_V$,
 $\omega_A$ and $\omega$ is to impose parity conservation for
 suitable matrix elements, for instance with the pseudoscalar density
 $P^{\pm}_x=\bar\chi_x\frac{\tau^\pm}{2}\gamma_5\chi_x$:
\be\label{eq2:14}
\sum_{\vec{x},\vec{y}} \langle \hat{A}^+_{x0}\, \hat{V}^-_{y0}\rangle
\;=\;
\sum_{\vec{x},\vec{y}} \langle \hat{V}^+_{x0}\, P^-_y\rangle\;=\; 0\ .
\ee
 These equations admit the solution
\begin{eqnarray}\label{eq2:15}
\tan{\omega}_V &=&
\frac{-i\sum_{\vec{x},\vec{y}}\langle V^+_{x0}\, P^-_y \rangle}
     {  \sum_{\vec{x},\vec{y}}\langle A^+_{x0}\, P^-_y \rangle} \ ,
\\[1em]\label{eq2:16}
\tan{\omega}_A &=&
\frac{i\!\sum_{\vec{x},\vec{y}}\langle A^+_{x0}\, V^-_{y0}\rangle
         \!+\!\tan{\omega}_V\!
         \sum_{\vec{x},\vec{y}}\langle A^+_{x0}\, A^-_{y0}\rangle} 
     {   \sum_{\vec{x},\vec{y}}\langle V^+_{x0}\, V^-_{y0}\rangle
         \!-\!i\!\tan{\omega}_V\!
         \sum_{\vec{x},\vec{y}}\langle V^+_{x0}\, A^-_{y0}\rangle} \ .
\end{eqnarray}

 In (\ref{eq2:14}) one can also take the derivatives of the currents
 instead of the currents themselves.
 For instance, taking the divergence of the vector current in the
 second equality gives the same equations as (\ref{eq2:13}) with
 $\hat{O}=P$.

 Once $\omega_V$ and $\omega_A$ are determined, the twist angle $\omega$
 can be obtained by eq.~(\ref{eq2:12}). 
 This method for determining the twist angle can also be used in case of
 simulations with partially quenched twisted mass quarks.
 The estimate of $\omega$ is, of course, affected by ${\cal O}(a)$
 ambiguities.
 For non-zero twisted mass $\mu\!\neq\!0$ the critical bare untwisted
 quark mass $\mu_\kappa=\mu_{\kappa cr}$, or the critical hopping
 parameter $\kappa_{cr}=(2\mu_{\kappa cr})^{-1}$, is signaled by
 $|\omega|\!=\!\pi/2$.

\subsection{Updating algorithm}\label{sec2.3}

 Concerning updating in our numerical simulations, we apply the two-step
 multi-boson (TSMB) algorithm \cite{TSMB}, which has been tuned to QCD
 simulations with Wilson quarks in previous works
 \cite{NF2TEST,VALENCE,SEA,SEAVAL,TMQCD}.
 (For details and references see in these papers.)
 In ref. \cite{NF2TEST} there is an approximate formula for the
 computational cost of an update cycle in terms of
 Matrix-Vector-Multiplications (MVMs):
\be \label{eq2:17}
N_{MVM}/\mbox{cycle} \;\simeq\;
6(n_B n_1 N_\Phi+N_U)+2n_B(n_2+n_3)N_{C}+I_G F_G \ .
\ee
 Here $n_{1,2,3}$ are the orders of polynomials used in the two
 approximation steps, $n_B$ gives the multiplicity in determinant
 breakup, $N_\Phi$ is the number of local bosonic sweeps per
 update cycle,  $N_U$ the number of local gauge sweeps, $N_{C}$ the
 number of global Metropolis accept-reject correction steps, and $I_G$ and
 $F_G$ give  the number of MVMs in the global boson heatbath and its
 frequency, respectively.
 
 The number of MVMs can also be converted into the number of floating
 point operations by noting that in our code, for vanishing twisted
 mass, we have
\be \label{eq2:18}
1\; {\rm MVM} \;\simeq\; 1.2\cdot 10^3\,\Omega\, {\rm flop} \ ,
\ee
 where $\Omega$ is the number of lattice points.
 For non-zero twisted mass there is an additional factor 2 due to the
 flavour index.
 (This does, however, not mean that twisted mass fermions are a factor
 of two more expensive since in this case the two flavours are
 incorporated in one fermion matrix and the polynomial approximations
 have lower orders: see appendix A.2 of \cite{TWIST}.)

 Measuring the integrated autocorrelations $\tau_{int}$ as a function
 of the quark mass in lattice units $am_q$ and of the lattice volume
 $\Omega$, previous experience tells that one can approximate the
 computational cost of a number of update cycles equal to $\tau_{int}$
 by the simple formula
\be\label{eq2:19}
C_{\tau_{int}} \simeq F\, (am_q)^{-z}\, \Omega \ .
\ee
 According to ref.~\cite{NF2TEST}, in case of combining the Wilson
 fermion action with the Wilson plaquette gauge action, the power of the
 inverse quark mass is close to $z=2$.
 The overall factor $F$ depends on the quantity under investigation.
 For Wilson quarks with Wilson gauge action the previous results
 can be summarized, for instance, for the average plaquette and for the
 pion mass determined with a randomly chosen source by
 \cite{SEAVAL}
\be\label{eq2:20}
F_{plaq} \simeq 7\cdot 10^6\, {\rm flop} \ , \hspace{3em}
F_{m_\pi} \simeq 10^6 \, {\rm flop} \ .
\ee
 Let us note that the approximate formula in (\ref{eq2:19}) has been, up
 to now, verified  only for some fixed values of the gauge coupling
 $\beta$.
 The $\beta$-dependence of $F$ has not yet been systematically
 investigated.

\section{Numerical simulation results}\label{sec3}

 Our aim is to compare the phase structure of two-flavour ($N_f=2$) QCD
 near zero quark mass for Wilson lattice fermion action and DBW2 gauge
 action with the one observed in ref.~\cite{TWIST,FERMILPROC} for Wilson
 fermion action and Wilson (plaquette-) gauge action.
 Since the phase structure obviously depends on the lattice spacing, we
 have to find the values of the bare parameters ($\beta,\mu_\kappa$) in
 the lattice action (\ref{eq2:01})-(\ref{eq2:03}) which correspond to
 quark mass $m_q \simeq 0$ and to the same lattice spacing as in
 \cite{TWIST,FERMILPROC}, namely $a \simeq 0.2\,{\rm fm}$.
 For having a fair comparison, the lattice volume has to be kept
 constant, too, because the metastability phenomenon does also depend
 on it.
 Therefore, we shall compare the results on $12^3 \times 24$ lattices.

 A possibility for facilitating the parameter tuning is to explore the
 position of the high-temperature phase transition on lattices with time
 extension $N_t=4$ and $N_t=6$ for small quark masses, which mark
 $a = 0.25-0.30\,{\rm fm}$ and $a = 0.17-0.20\,{\rm fm}$, respectively.
 (This method with $N_t=4$ has been applied, for instance, in
 \cite{NF2TEST}.)
 A useful first orientation is also provided by the quenched studies.
 (For a useful collection of data on RGI gauge actions see \cite{NECCO}
 and references therein).
 For specifying the actual value of the lattice spacing we
 determine the Sommer scale parameter in lattice units $r_0/a$
 \cite{SOMMER}, which we set by definition to be
 $r_0 \equiv 0.5\,{\rm fm}$, independently from the quark mass.

\begin{figure}
\begin{center}
\begin{minipage}[c]{1.0\linewidth}
 \begin{minipage}[c]{0.65\hsize}
  \includegraphics[angle=-90,width=.95\hsize]
   {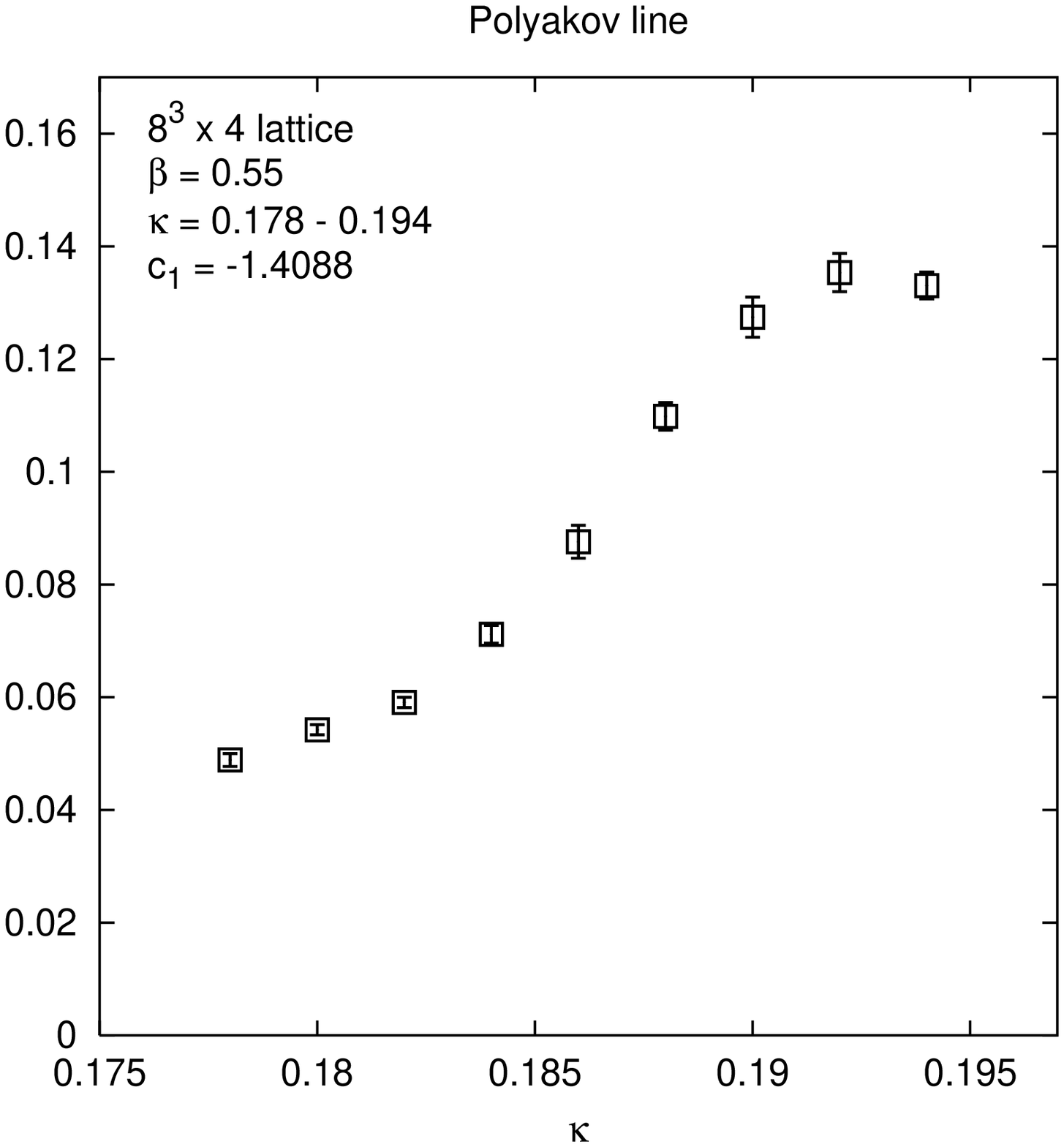}
 \end{minipage}
 \hspace*{-25mm}
 \begin{minipage}[c]{0.65\hsize}
  \includegraphics[angle=-90,width=.95\hsize]
   {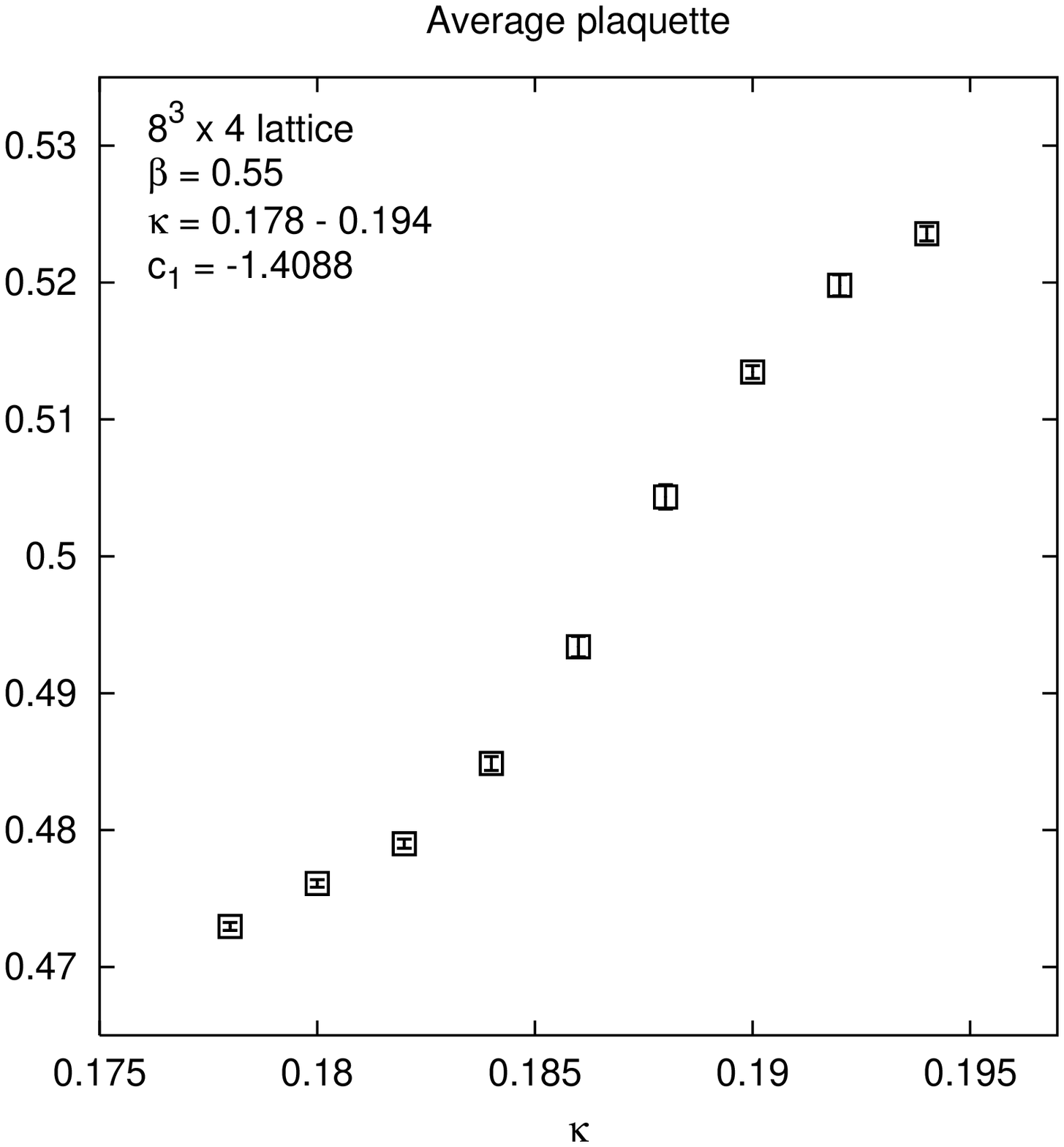}
 \end{minipage}
\end{minipage}
\end{center}
\vspace*{2mm}
\begin{center}
\begin{minipage}[c]{1.0\linewidth}
 \begin{minipage}[c]{0.65\hsize}
  \includegraphics[angle=-90,width=.95\hsize]
   {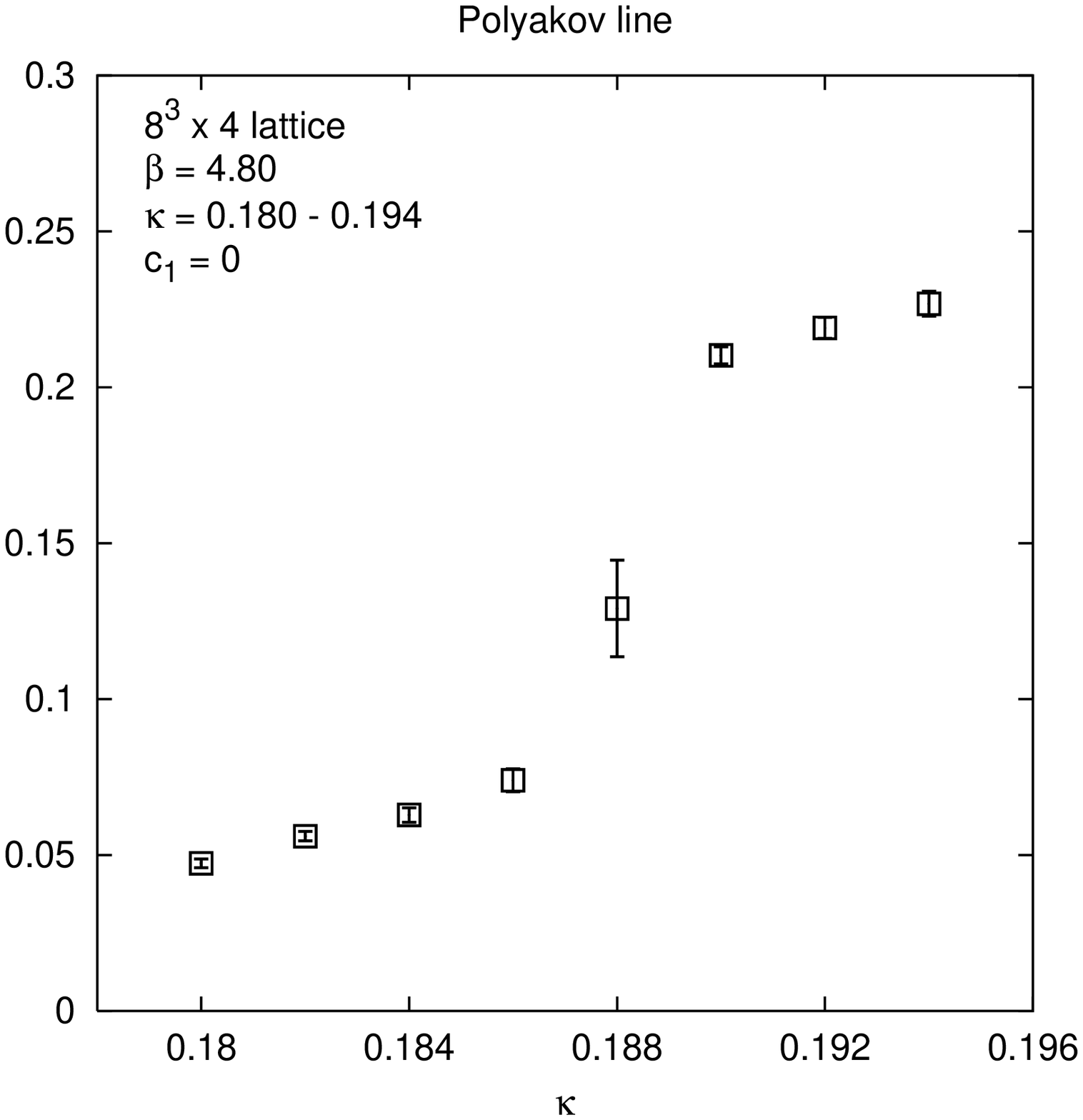}
 \end{minipage}
 \hspace*{-25mm}
 \begin{minipage}[c]{0.65\hsize}
  \includegraphics[angle=-90,width=.95\hsize]
   {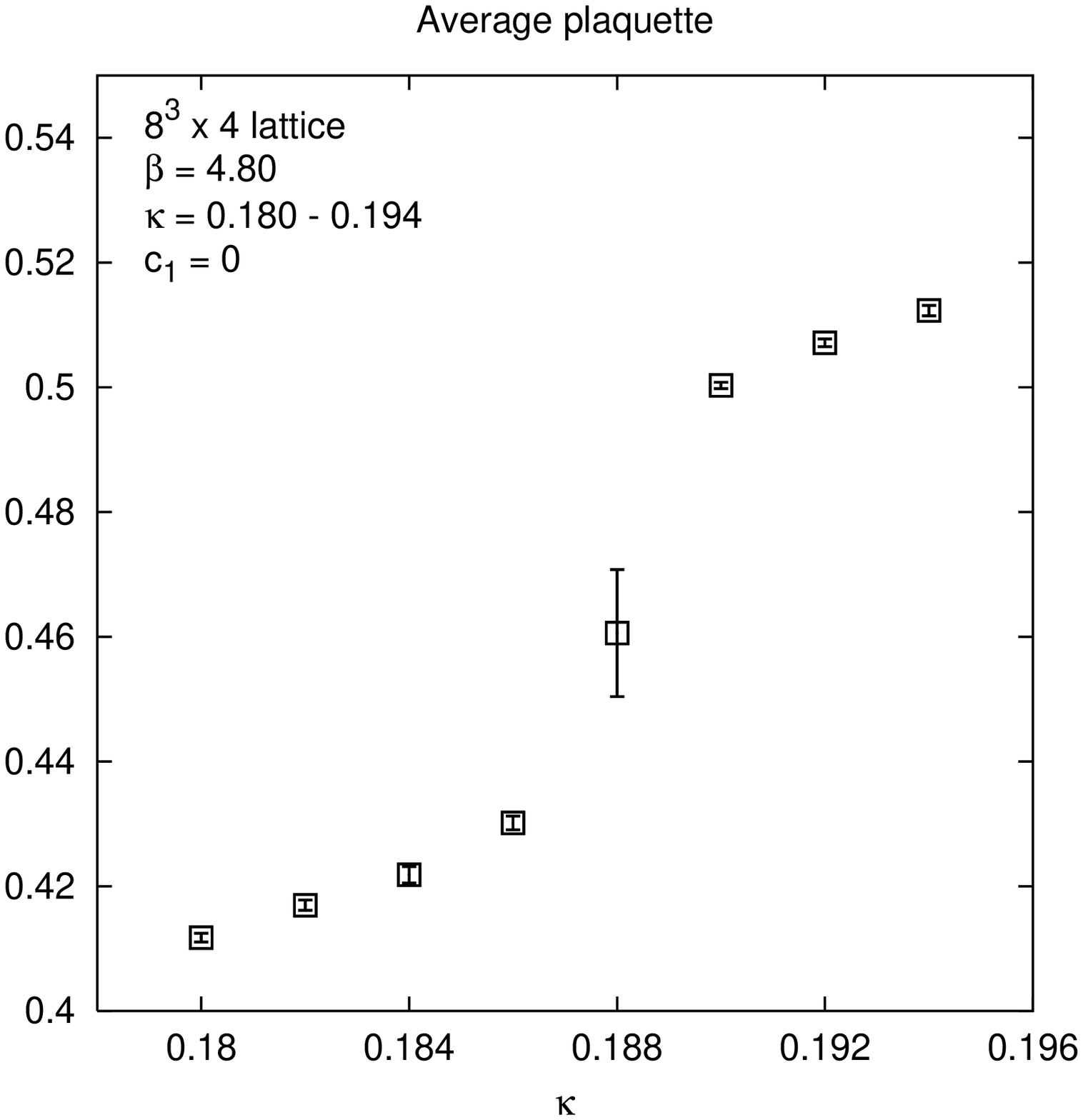}
 \end{minipage}
\end{minipage}
\end{center}
\vspace*{2mm}
\begin{center}
\parbox{0.8\linewidth}{\caption{\label{fig_Nt=4}\em
 Upper panels: the signals of the $N_t=4$ non-zero temperature
 transition on an $8^3\times 4$ lattice with the DBW2 gauge action.
 Lower panels: the same with Wilson gauge action.
 Left panels: absolute value of the Polyakov line,
 right panels: average Wilson loop, both as a function of $\kappa$.}}
\end{center}
\end{figure}

 In order to localize the $N_t=4$ high-temperature phase transition
 we fixed the gauge coupling at $\beta=0.55$ and changed the bare quark
 mass $\mu_\kappa$ (or, equivalently, the hopping parameter
 $\kappa=(2\mu_\kappa)^{-1}$).
 The results on an $8^3 \times 4$ lattice for the absolute value of the
 Polyakov-line and average plaquette are given in figure~\ref{fig_Nt=4}.
 As it is shown by the figure, the transition with the DBW2 action is
 rather smooth, barely visible.
 This has to be contrasted with the strong and sudden increase of both
 Polyakov-line and average plaquette in case of the Wilson plaquette
 action, which is also shown for comparison in figure~\ref{fig_Nt=4}.

 A similar analysis on $12^3 \times 6$ lattices at $\beta=0.67$ gives
 qualitatively similar results but there the difference between the DBW2
 and the Wilson plaquette action is smaller because the transition
 for the Wilson plaquette action becomes weaker.

\subsection{Results on an $8^3 \times 16$ lattice at $\beta=0.55$}
\label{sec3.1}

 The runs on an $8^3 \times 16$ lattice at $\beta=0.55$ and $\mu=0$ were
 started from the low-temperature phase by taking four copies in the
 time direction of some of the $8^3 \times 4$ lattices.
 The parameters of these runs are specified in the first part of
 table~\ref{tab_run}.
 Besides the hopping parameter $\kappa$ also some parameters of the
 TSMB updating algorithm are specified: the orders of the polynomials
 used $n_{1,2,3}$ and the interval covering the eigenvalues of the
 squared preconditioned hermitean quark matrix $[\epsilon,\lambda]$.

\begin{figure}
\vspace*{-0.03\hsize}
\begin{center}
\begin{minipage}[c]{0.85\linewidth}
 \hspace*{0.10\hsize}
 \begin{minipage}[c]{0.95\hsize}
  \includegraphics[angle=-90,width=1.\hsize]
   {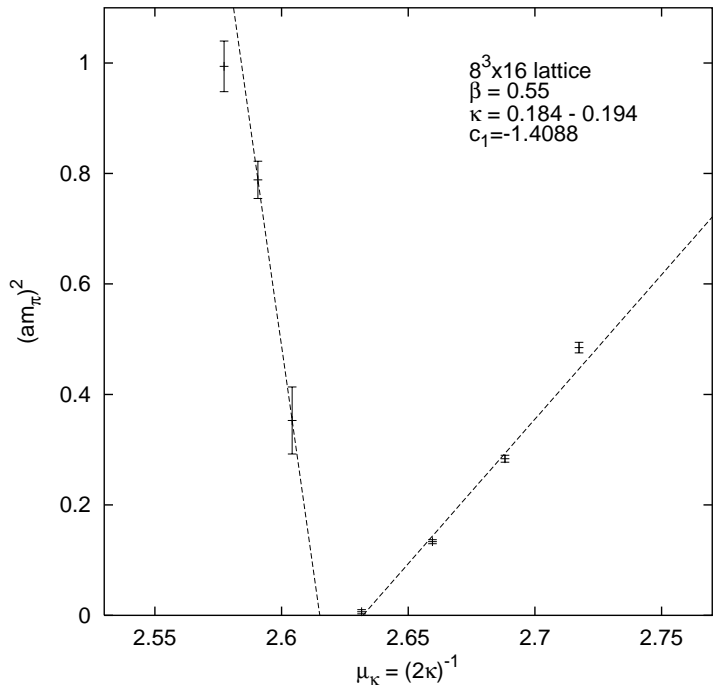}
 \vspace*{0.03\hsize}
 \end{minipage}
 \hspace*{0.075\hsize}
 \begin{minipage}[c]{0.95\hsize}
  \includegraphics[angle=-90,width=1.\hsize]
   {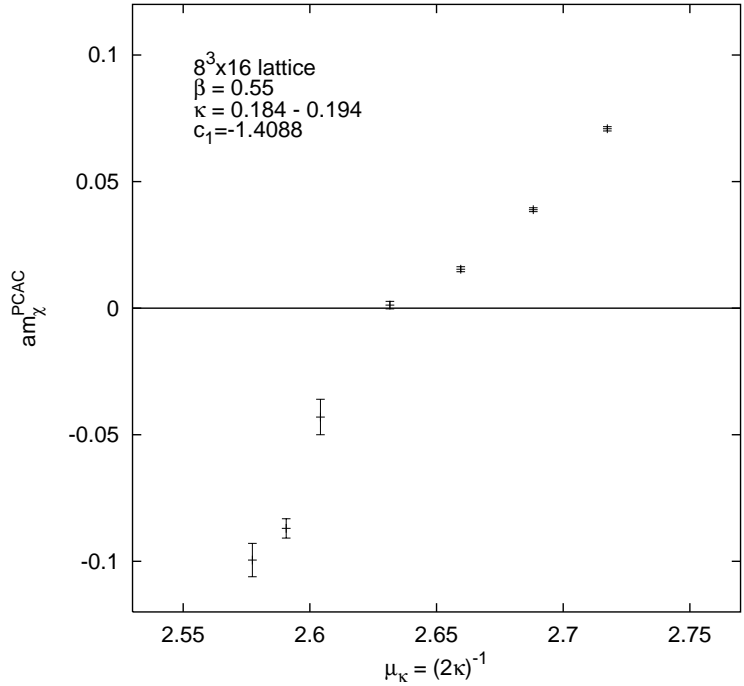}
 \end{minipage}
\end{minipage}
\end{center}
\vspace*{-0.04\hsize}
\begin{center}
\parbox{0.8\linewidth}{\caption{\label{fig_AMpisq_Zqamq_8c16}\em
 Results of the numerical simulation on an $8^3\times 16$ lattice at
 $\beta=0.55$: upper panel the square of the pion mass $(am_\pi)^2$,
 lower panel the PCAC quark mass $am_\chi^{PCAC}$.
 In the upper panel the dashed lines are extrapolations to zero pion
 mass: at right it is a linear fit of four points, at left a straight
 line connecting two points with small quark mass.}}
\end{center}
\end{figure}

 In the $8^3\times 16$ runs we looked for signals of metastability but
 we did not find any.
 The results for some interesting quantities are collected in the first
 part of table~\ref{tab_resu}: the pion (i.e.~pseudoscalar meson) and
 $\rho$-meson masses and the bare quark mass in lattice units
 $am_\chi^{PCAC}$.
 Some of these quantities are also shown in
 figure~\ref{fig_AMpisq_Zqamq_8c16}.
 The scale parameter in lattice units $r_0/a$ was also determined.
 We note in passing that at this small value of $\beta$ and with our
 partly low statistics the evaluation of $r_0$ is rather difficult. 
 Nevertheless, in order to estimate quantities also in physical units, 
 we performed a purely statistical analysis for $r_0$, being aware of
 the fact that systematic effects can be large.

 The bare quark mass $am_\chi^{PCAC}$ is defined by the PCAC-relation
 containing the axialvector current $A^a_{x\mu}$ in (\ref{eq2:05}) and
 the pseudoscalar density insertion:
\be\label{eq3:01}
am_\chi^{PCAC} \equiv 
\frac{\langle \partial^\ast_\mu A^+_{x\mu}\, P^-_y \rangle}
{2\langle P^+_x\, P^-_y \rangle} \ .
\ee
 Since for the moment we do not determine the $Z$-factors of
 multiplicative renormalization, the bare quark mass $am_\chi^{PCAC}$
 contains an unknown ${\cal O}(1)$ $Z$-factor $Z_q \equiv Z_P/Z_A$.
 In the following analysis we extracted the quark mass with the method 
 detailed in ref.~\cite{NF2TEST}, see section 3.1.1 there.

 In agreement with the absence of a signal for metastability, the
 $\mu_\kappa$-dependence of the pion mass and quark mass in
 figure~\ref{fig_AMpisq_Zqamq_8c16} is consistent with the absence of
 a first order phase transition at this gauge coupling ($\beta=0.55$).
 A rough estimate for the value of the lattice spacing is
 $a \simeq 0.30\,{\rm fm}$ in the positive quark mass phase and
 $a \simeq 0.23\,{\rm fm}$ in the negative quark mass phase.
 The upper panel in figure~\ref{fig_AMpisq_Zqamq_8c16} suggests the
 existence of a short interval ($\mu_\kappa \in [2.62,2.63]$ or
 $\kappa \in [0.190,0.191]$) of an Aoki phase near zero quark- and
 pion-masses.
 This behaviour is qualitatively similar to the one for the Wilson
 plaquette action which also shows the existence of the Aoki phase
 at strong gauge coupling \cite{HUMBOLDT}.

\subsection{Results on a $12^3 \times 24$ lattice at $\beta=0.67$}
\label{sec3.2}

 With a short investigation of the high-temperature phase transition on
 a $12^3\times 6$ lattice one can easily localize the gauge coupling
 $\beta$ and bare quark mass $\mu_\kappa=(2\kappa)^{-1}$ where the lattice
 spacing is about a factor 3/2 smaller than at $\beta=0.55$.
 It turned out that one can take $\beta=0.67$ and $\kappa\simeq 0.17$.
 Fixing $\beta=0.67$ and changing $\kappa$ we performed several runs
 on a $12^3\times 24$ lattice.
 In this way the physical volume of the lattice is approximately the
 same as the one of an $8^3\times 16$ lattice at $\beta=0.55$.
 In order to be able to compare with the results of ref.~\cite{TWIST},
 besides the runs with $\mu=0$, at this $\beta$ we also considered a
 non-vanishing twisted mass $\mu=0.01$.

\begin{figure}
\vspace*{-0.04\hsize}
\begin{center}
\hspace*{0.15\linewidth}
\begin{minipage}[c]{1.0\linewidth}
\includegraphics[angle=-90,width=.90\hsize]
 {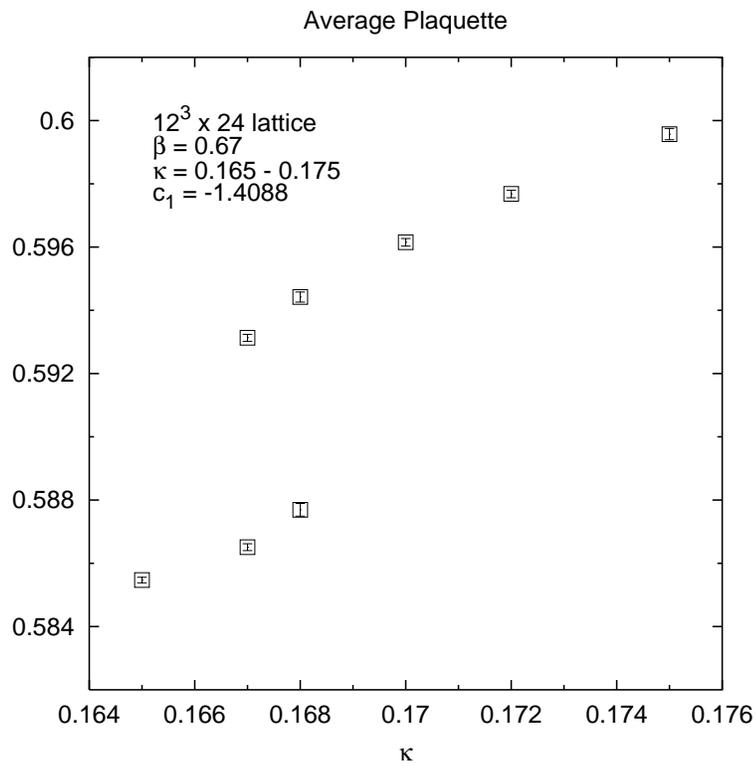}
\end{minipage}
\end{center}
\vspace*{-0.03\hsize}
\begin{center}
\hspace*{0.15\linewidth}
\begin{minipage}[c]{1.0\linewidth}
\includegraphics[angle=-90,width=.90\hsize]
 {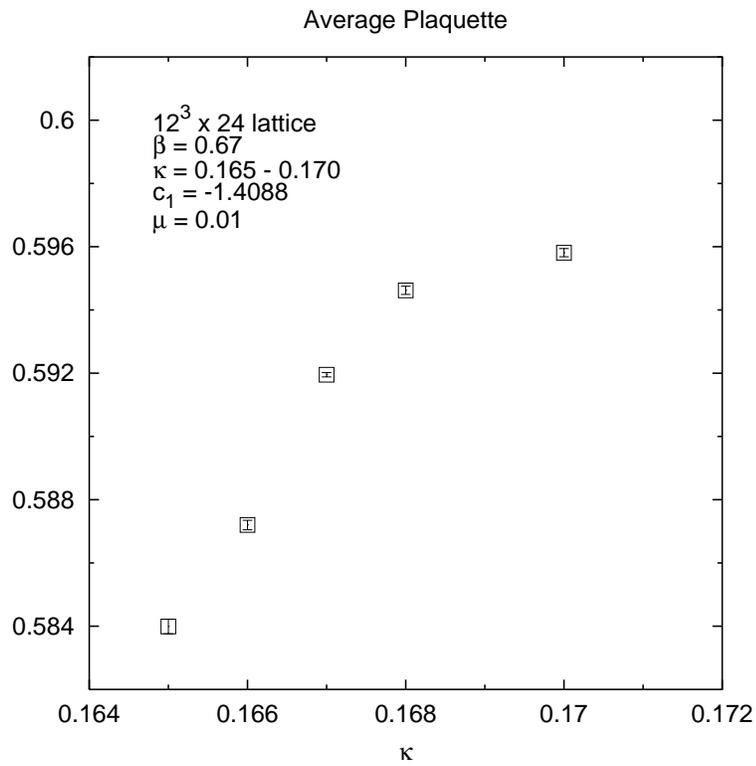}
\end{minipage}
\end{center}
\vspace*{-0.05\hsize}
\begin{center}
\parbox{0.8\linewidth}{\caption{\label{fig_Aveplaq}\em
 The average plaquette at $\beta=0.67$ on a $12^3\times 24$ lattice as
 a function of the hopping parameter $\kappa$:
 upper panel $\mu=0$ and lower panel $\mu=0.01$, respectively.}}
\end{center}
\end{figure}

 First we looked also here at $\mu=0$ for a signal of metastability in
 the average plaquette and we found it near $\kappa=0.167-0.168$, as it
 is shown by the upper panel of figure~\ref{fig_Aveplaq}.
 Note that the average plaquette values are substantially higher here
 than at $\beta=5.2$ with the Wilson plaquette gauge action in
 \cite{TWIST}: $A_{plaq} \equiv
 \langle\frac{1}{3}{\rm Re\,Tr\,}U_{plaq}\rangle \simeq 0.59$ instead
 of $A_{plaq} \simeq 0.52$.
 This qualitatively shows that the gauge field with DBW2 is smoother.

\begin{figure}
\begin{center}
\begin{minipage}[c]{1.15\linewidth}
 \begin{flushleft}
  \begin{minipage}[c]{0.57\hsize}
   \includegraphics[angle=-90,width=1.\hsize]
    {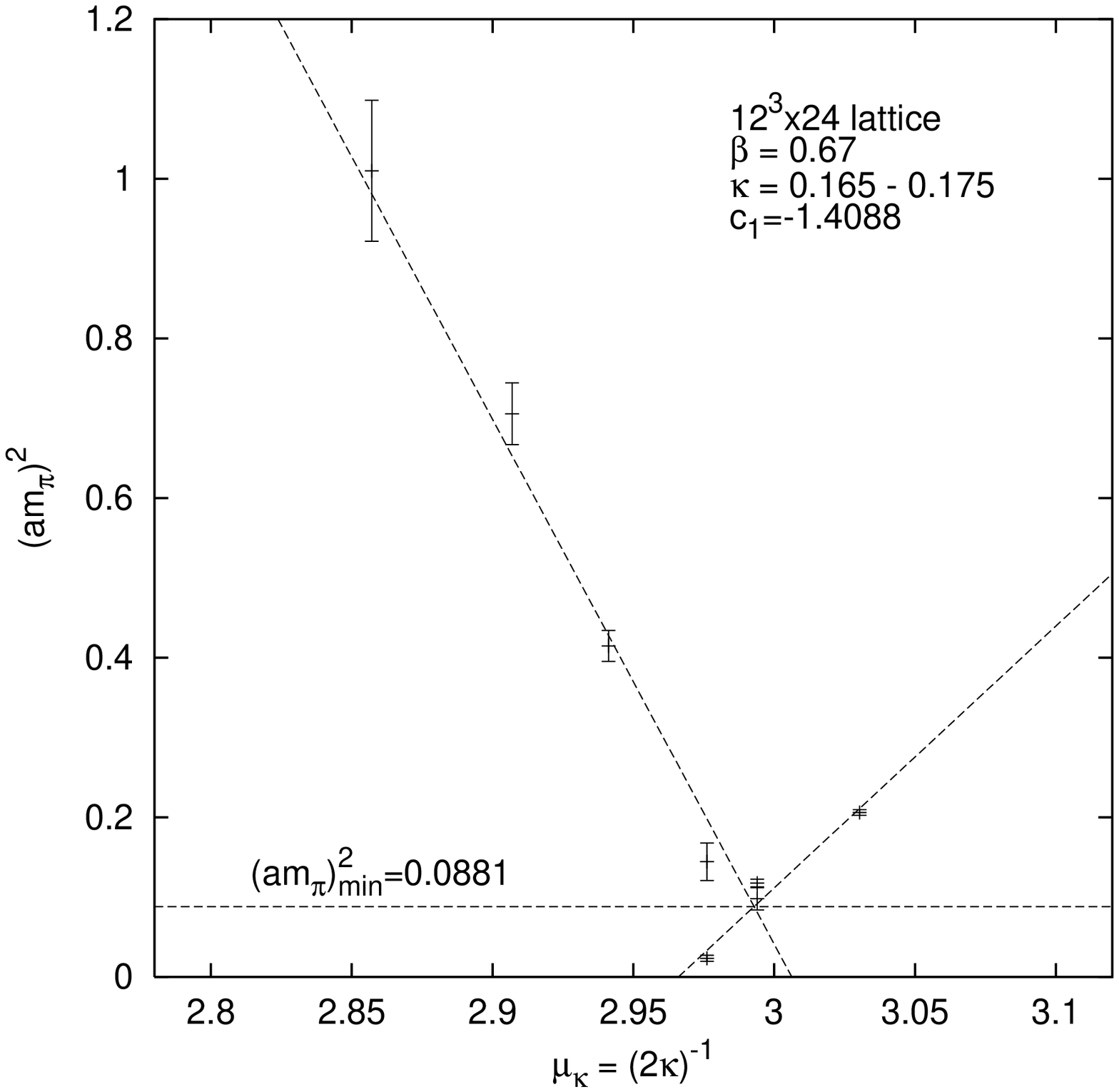}
  \end{minipage}
 \end{flushleft}
 \vspace*{-0.445\hsize}
 \begin{flushright}
  \begin{minipage}[c]{0.57\hsize}
   \includegraphics[angle=-90,width=1.\hsize]
    {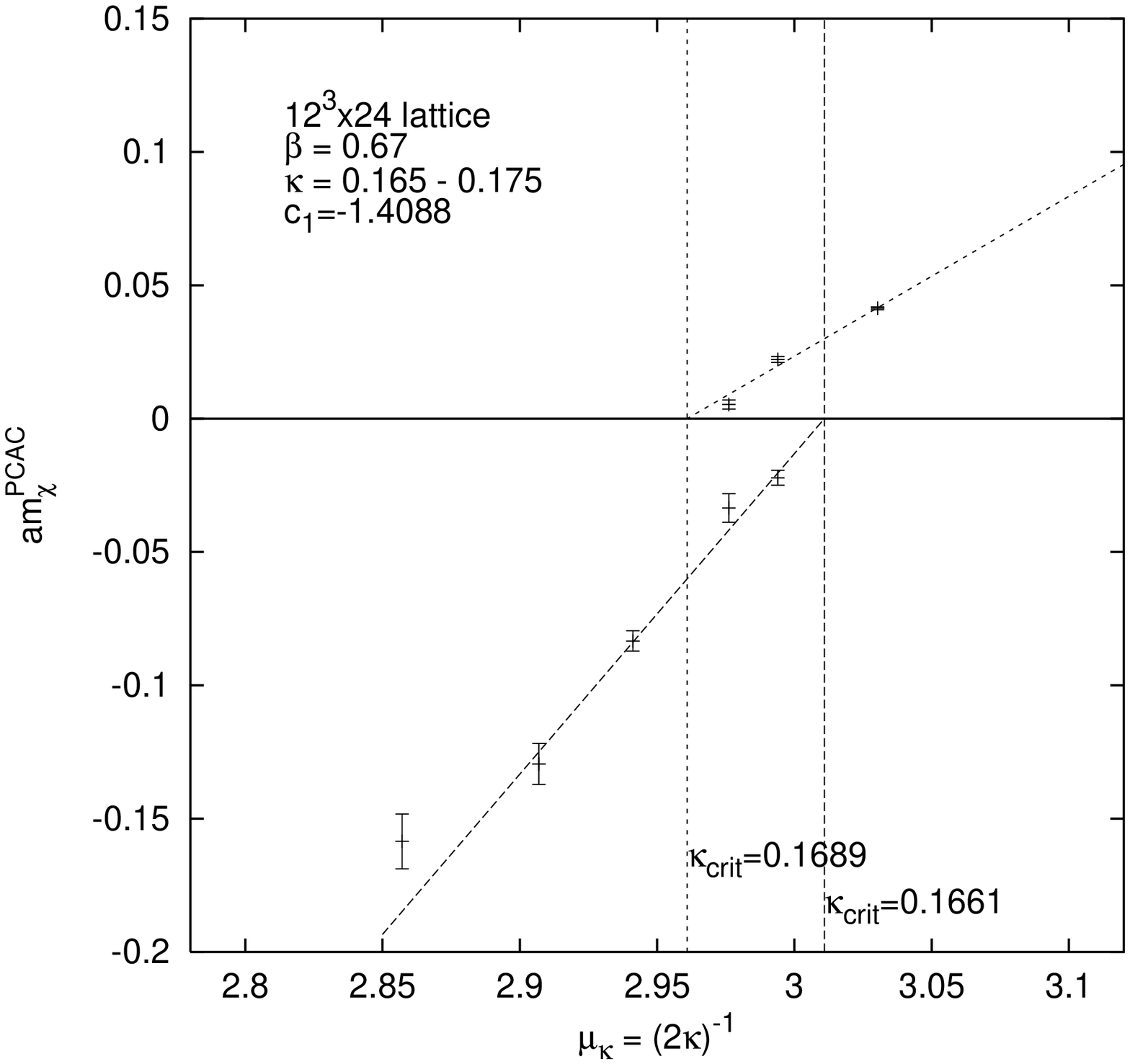}
  \end{minipage}
 \end{flushright}
 \vspace*{0.01\hsize}
  \begin{flushleft}
  \begin{minipage}[c]{0.57\hsize}
   \includegraphics[angle=-90,width=1.\hsize]
    {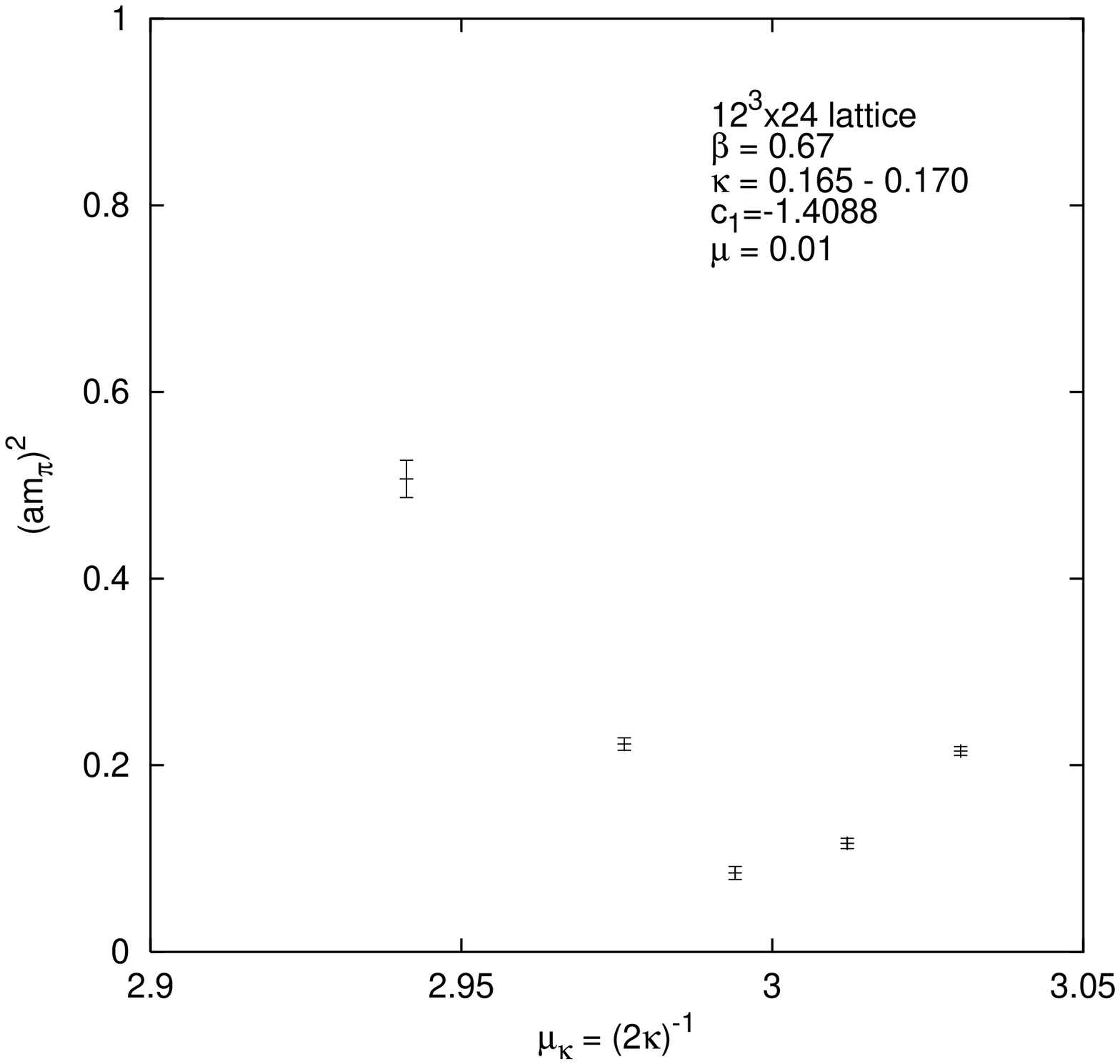}
  \end{minipage}
 \end{flushleft}
 \vspace*{-0.445\hsize}
 \begin{flushright}
  \begin{minipage}[c]{0.57\hsize}
   \includegraphics[angle=-90,width=1.\hsize]
    {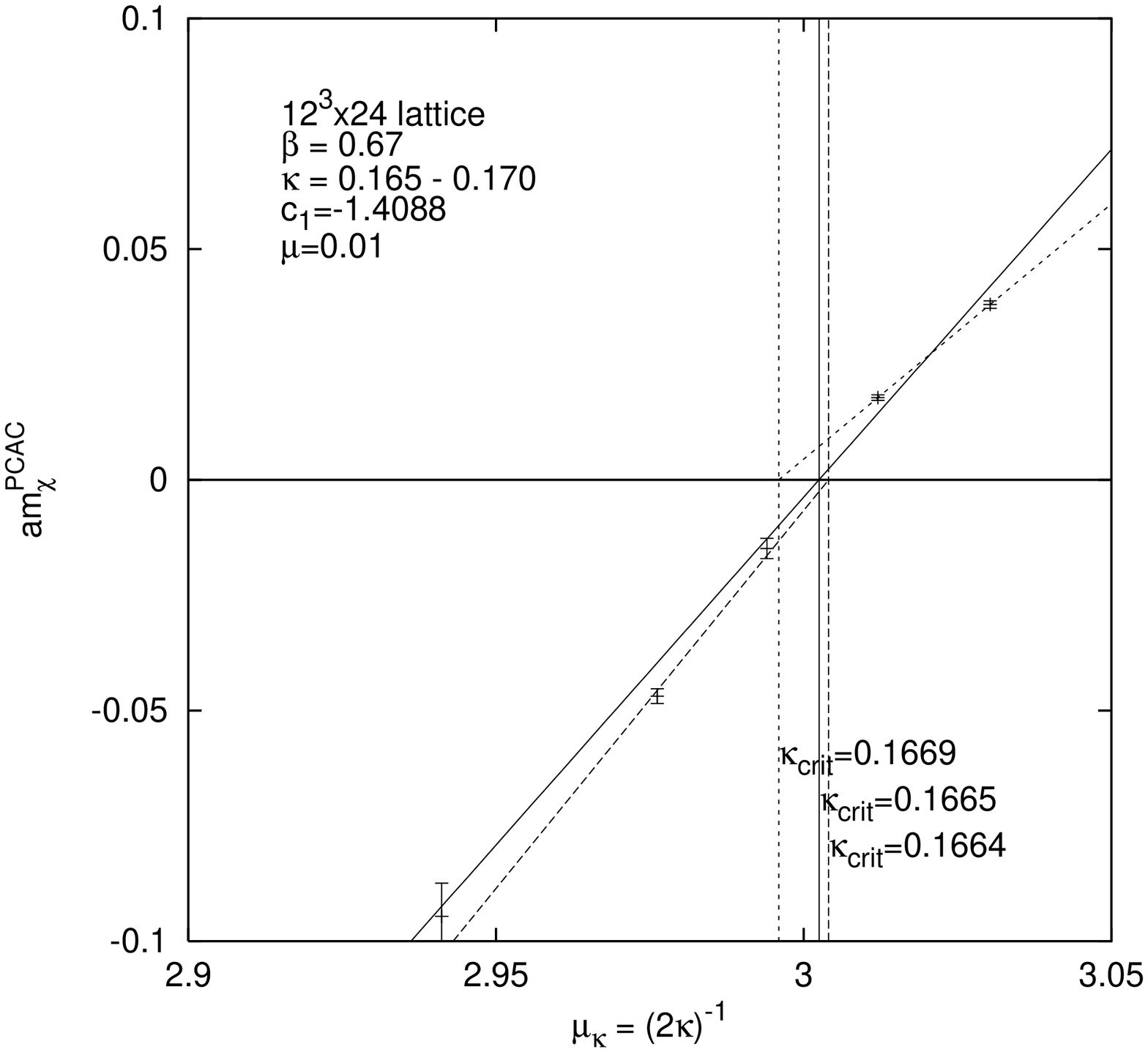}
  \end{minipage}
 \end{flushright}
\end{minipage}
\end{center}
\vspace*{-0.01\hsize}
\begin{center}
\parbox{0.8\linewidth}{\caption{\label{fig_AMpisq_Zqamq_12c24}\em
 Results of the numerical simulation on a $12^3\times 24$ lattice at
 $\beta=0.67$ as a function of $\mu_\kappa=(2\kappa)^{-1}$: upper panels
 $\mu=0$, lower panels $\mu=0.01$.
 Left panels: $(a m_\pi)^2$, right panels: the bare PCAC quark mass
 $am_\chi^{PCAC}$.
 The straight lines are fits to the points in the positive and negative
 quark mass phase, respectively.
 The horizontal line in the upper left panel shows the estimated value
 of the minimal pion mass in lattice units.
 The straight lines in the right panels are explained in the text.}}
\end{center}
\end{figure}

 We also determined the pion, $\rho$-meson and quark masses, with the
 results given in table~\ref{tab_resu}.
 (For a graphical representation of some of these results see also the
 upper panels of figure~\ref{fig_AMpisq_Zqamq_12c24}.)
 For the extraction of $r_0/a$ we performed only a statistical
 analysis also here, neglecting the systematic effects.
 Let us give a range of values for the runs in table~\ref{tab_resu}. 
 For $A_{\mbox{l}}$ to $D_{\mbox{l}}$ we find $2.37 < r_0/a < 2.76$ in
 the low and for $C_{\mbox{h}}$ to $G_{\mbox{h}}$
 $2.72 <r_0/a < 3.17$ in the high plaquette phases, respectively.
 For $\mathcal{A}_{\mbox{l}}, \mathcal{B}_{\mbox{l}}$ we find
 $2.39 <r_0/a < 2.54$ in the low and for $\mathcal{C}_{\mbox{h}}$ to
 $\mathcal{E}_{\mbox{h}}$ we find $2.89 <r_0/a < 3.07$ in the high
 plaquette phase.

 From the values of the scale parameter $r_0/a$ we determined the
 lattice spacing, and found $a \simeq 0.18-0.21\,{\rm fm}$ in the
 positive and $a \simeq 0.16-0.18\,{\rm fm}$ in the negative quark mass
 phase, respectively.
 This is quite close to the values obtained in both phases with the
 Wilson plaquette gauge action at $\beta=5.2$ in ref.~\cite{TWIST}.

 Going to the positive twisted mass $\mu=0.01$, the metastability in the
 average plaquette disappears on our $12^3\times 24$ lattice, as it is
 shown by the lower panel of figure~\ref{fig_Aveplaq}.
 Having in mind the strong metastability signal in the average plaquette
 observed on a $12^3\times 24$ lattice at $\beta=5.2$ and $\mu=0.01$ with
 the Wilson plaquette gauge action in \cite{TWIST}, the absence of the
 metastability here signals a dramatic improvement of the phase structure
 due to the DBW2 gauge action.
 The presence of metastability at $\mu=0$ and the absence of it at
 $\mu=0.01$ indicates the existence of a rather short first order phase
 transition line near the origin in the $(\mu_\kappa,\mu)$-plane.
 Of course, for a precise localization of the first order phase
 transition line a detailed study of the infinite volume limit is
 required, which is beyond the scope of this paper.
 
 An important question is the minimal value of the pion mass
 $m_\pi^{min}$ associated to the first order phase transition line.
 A precise definition of $m_\pi^{min}$ could be the value of the
 infinite volume pion mass just at the position of the first order phase
 transition, defined by the equal depth of the two free energy minima in
 the infinite volume limit.
 To obtain this would be rather demanding.
 Although the volume dependence could be studied beyond our volume
 extension of $L \simeq 2.4\,{\rm fm}$, for instance on a $16^3\times 32$
 lattice, the precise comparison of the free energy minima would be
 quite difficult.
 An approximate determination of $m_\pi^{min}$ can be obtained by
 requiring the equality of the pion mass in lattice units $a m_\pi$ in
 the two phases on our $12^3\times 24$ lattices.
 For this a linear extrapolation of $(a m_\pi)^2$ from the points on
 both sides of the phase transition can be considered.
 As shown in the upper left panel of
 figure~\ref{fig_AMpisq_Zqamq_12c24}, our result at $\mu=0$ is
 $(am_\pi)^2=0.0881$.
 This implies, with the range of $r_0$ values given above, that
 $m_\pi^{min} \simeq 251\,{\rm MeV}$ in the positive
 quark mass phase and $m_\pi^{min} \simeq 374\,{\rm MeV}$ in the phase
 with negative quark mass.

 The minimal charged pion mass at $\mu=0.01$ is
 $m_\pi^{min} \simeq 360\,{\rm MeV}$ (see lower left panel of
 figure~\ref{fig_AMpisq_Zqamq_12c24}).
 This originates from the non-zero value of the twisted quark mass and
 not from the presence of a first order phase transition.

 In the right panels of figure~\ref{fig_AMpisq_Zqamq_12c24} the bare
 quark mass in lattice units is shown.
 The dashed lines are linear fits to the points with positive and
 negative quark mass, respectively.
 At zero twisted mass (upper panel) the metastability region near the
 first order phase transition is clearly displayed.
 At $\mu=0.01$ (lower panel) the difference between the two dashed lines
 is smaller.
 This difference may be interpreted as a consequence of a cross-over in
 the continuation of the first order phase transition line.
 In the figure there is also a linear fit to all points shown (full
 line) which goes reasonably close to every point.
 The two dashed lines also give lower and upper bound estimates for the
 critical hopping parameter: $0.1661 \leq \kappa_{cr} \leq 0.1689$.

 Another way to estimate the critical hopping parameter (i.e. critical
 bare untwisted quark mass) is to determine the twist angle and find
 $\kappa_{cr} = (2\mu_{\kappa cr})^{-1}$ where it equals $\pi/2$.
 Considering, for definiteness, the twist angle $\omega_V$ defined
 in section \ref{sec2.2}, the fit in figure~\ref{fig_omegav} gives
 $\kappa_{cr} = 0.16651(2)$, in good agreement with the previous
 estimate.
 (Actually the numbers in figure~\ref{fig_omegav} come from the
 vector Ward identity (\ref{eq2:13}) but, within errors,
 (\ref{eq2:15}) gives compatible results.)
 The $Z$-parameter appearing in this fit for $\omega_V$ is
 $Z_{oV} \equiv Z_A Z_S/(Z_V Z_P)$ (see section \ref{sec2.2} and
 \cite{ALPHA-TMQCD}).
 According to figure~\ref{fig_omegav} we have $Z_{oV} = 0.959(30)$.
 Since from an analogous fit to $\omega_A$ one could determine
 $Z_{oA} \equiv Z_V Z_S/(Z_A Z_P)$, this also offers a relatively easy
 way to obtain the $Z$-parameter combinations $Z_A/Z_V$ and $Z_P/Z_S$.

\begin{figure}
\begin{center}
\hspace*{0.08\linewidth}
\begin{minipage}[c]{1.2\linewidth}
\includegraphics[angle=-90,width=.95\hsize]
 {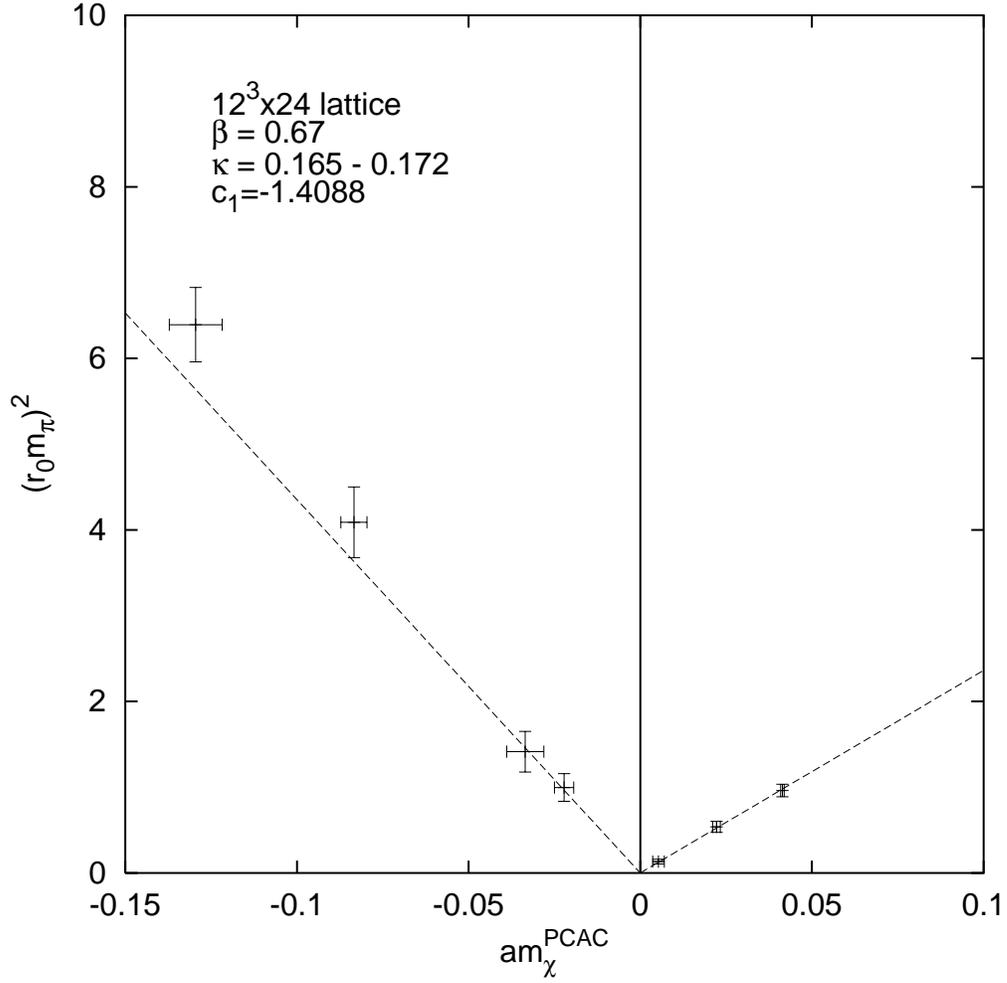}
\end{minipage}
\end{center}
\begin{center}
\parbox{0.8\linewidth}{\caption{\label{fig_MrZqamq}\em
 The dependence of $(r_0 m_\pi)^2$ on $am_\chi^{PCAC}$ at
 $\beta=0.67$, $\mu=0$ on a $12^3\times 24$ lattice.
 The dashed lines are linear fits of the form $A|am_\chi^{PCAC}|$.
 At right all three points are included in the fit, at left only
 two of them with the smallest quark masses.
 The fit parameter is for positive and negative quark mass
 $A=23.6$ and $A=43.5$, respectively.}}
\end{center}
\end{figure}

\begin{figure}
\begin{center}
\hspace*{0.08\linewidth}
\begin{minipage}[c]{1.2\linewidth}
\includegraphics[angle=-90,width=.95\hsize]
 {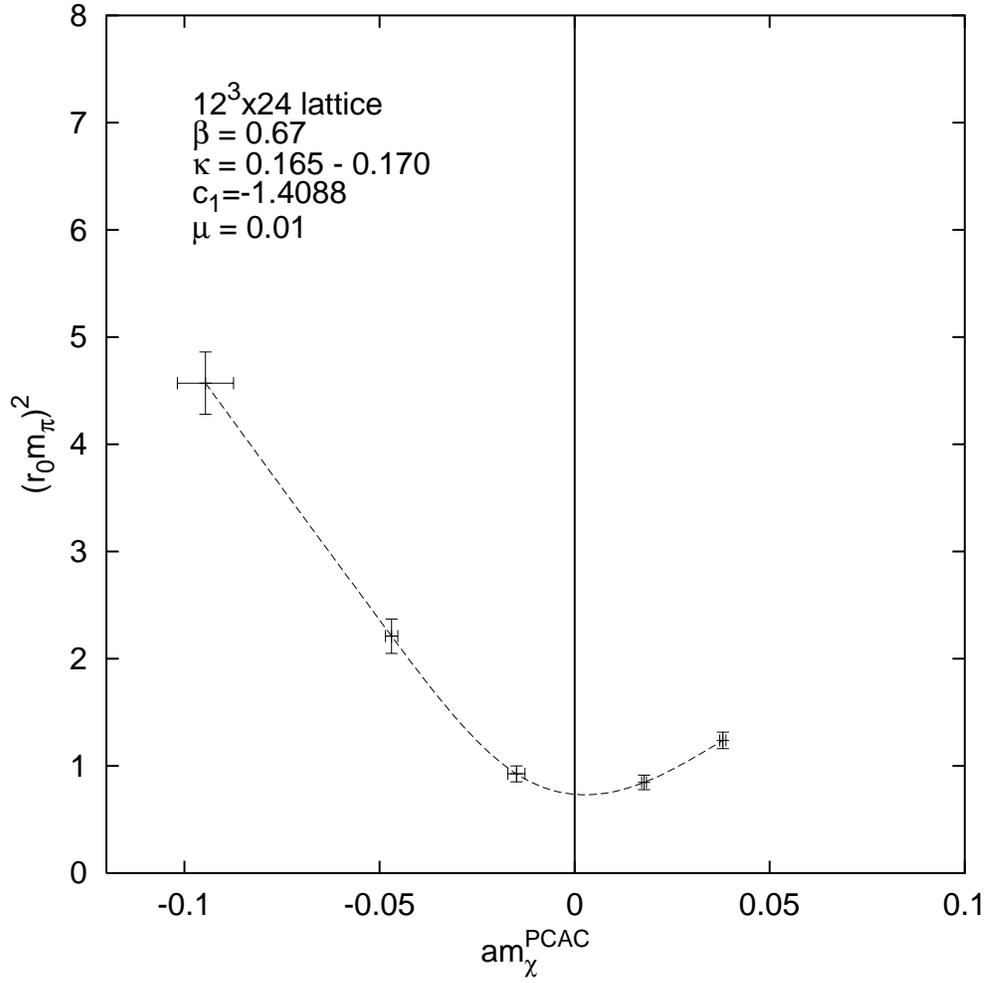}
\end{minipage}
\end{center}
\begin{center}
\parbox{0.8\linewidth}{\caption{\label{fig_MrZqamq_TM}\em
 The dependence of $(r_0 m_\pi)^2$ on $am_\chi^{PCAC}$ at
 $\beta=0.67$, $\mu=0.01$ on a $12^3\times 24$ lattice.
 The dashed line is a spline interpolation for guiding the eyes.}}
\end{center}
\end{figure}

 The quantities $(r_0 m_\pi)^2$ and $am_\chi^{PCAC}$ can also be
 plotted against each other (see figure~\ref{fig_MrZqamq} and
 \ref{fig_MrZqamq_TM} for $\mu=0$ and $\mu=0.01$, respectively).
 Figure~\ref{fig_MrZqamq} and the data in table~\ref{tab_resu} show
 that at $\mu=0$, in the metastable region beyond the minimal pion mass,
 one can also reach values close to the physical value
 $m_\pi \simeq 140\,{\rm MeV}$.

\subsection{Topological charge}
\label{sec3.3}

 The RGI gauge actions, and in particular the DBW2 action, are known
 to slow down the transitions between different topological sectors
 both in quenched \cite{RBC-PHYSREV} and in dynamical domain wall
 simulations \cite{DYNAMICALDW}.
 In order to check this we determined the topological charge $Q_{top}$
 in some of the runs using a cooling method \cite{COOLING}.
 In the following we denote the result from the cooling analysis by
 ``topological charge'', being aware of the fact that this definition
 contains some degree of arbitrariness.
 However, for our aim of testing the autocorrelation time this
 definition is sufficient.

 In the run with label $(\mathcal{C}_{\mbox{h}})$ ($12^3\times 24\;
 {\rm lattice},\; \beta=0.67,\; \mu=0.01,\; \kappa=0.167$) the history
 of the topological charge is shown in the upper panel of
 figure~\ref{fig_Qhist}.
 (The lower panel is a histogram of $Q_{top}$.)
 The analyzed configurations are separated by 10 TSMB update cycles.
 In this point, according to table~\ref{tab_resu}, the quark mass is
 about $m_q \simeq 0.3\, m_{\rm strange}$ and the pion mass
 $m_\pi \simeq 380\,{\rm MeV}$.
 As the figure shows, the topological charge is often changed.
 Its integrated autocorrelation in this run is
 $\tau^{top}_{int} \simeq 180$, but there is obviously a long tail of
 the autocorrelation which is not yet properly taken into account in a
 run of this length. 
 In any case, $\tau^{top}_{int}$ is substantially longer than those of
 the average plaquette ($\tau^{plaq}_{int} \simeq 22$) or of the pion
 mass ($\tau^{m_\pi}_{int} \simeq 6$) in table~\ref{tab_auto}.

 In another run, the one with label $(C_l)$ ($12^3\times 24\;
 {\rm lattice},\; \beta=0.67,\; \mu=0,\; \kappa=0.167$), where
 the quark mass is about $m_q \simeq 0.18\, m_{\rm strange}$ and the
 pion mass $m_\pi \simeq 295\,{\rm MeV}$, the general picture is similar
 to figure~\ref{fig_Qhist}.
 The integrated autocorrelation here comes out to be
 $\tau^{top}_{int} \simeq 70$, but this value is even less reliable
 because the run is shorter.

 In spite of these relatively long autocorrelations, it is clear that
 in a sufficiently long run, say of length $1000\,\tau^{m_\pi}_{int}$,
 which would be needed anyway for a good statistics on other quantities,
 the different topological sectors could be properly sampled by the TSMB
 algorithm.
 Therefore, at these bare parameter values, there is no problem with the
 suppression of the transitions between different topological sectors.

\subsection{Results about the update algorithm}
\label{sec3.4}

 In this paper we applied the TSMB update algorithm \cite{TSMB}.
 The estimates of the autocorrelations in different runs and the cost
 estimates obtained using (\ref{eq2:17}) are given in
 table~\ref{tab_auto}.
 Since in our relatively short runs the autocorrelations can only be
 estimated, say, within a factor of two, the numbers in
 table~\ref{tab_auto} give only a first orientation.

 Qualitatively speaking, the $12^3\times 24$ runs with ``low-plaquette''
 (positive quark mass) have lower costs than the corresponding runs with
 ``high plaquette'' (negative quark mass): at the same absolute value of
 the bare quark mass the runs in the negative quark mass phase have 
 in most cases at least by an order of magnitude higher costs than those
 in the positive quark mass phase.
 The reason of the higher cost at negative quark mass is that the
 smallest eigenvalues fluctuate more frequently to very small values.

 There is also a general tendency that the overall factors $F$ decrease
 for decreasing absolute value of the quark mass.
 In fact, the data on $F$ show that in the small quark mass region an
 inverse quark mass power $z=1$ is a better approximation
 than $z=2$, which has been observed in previous simulations with the
 Wilson plaquette gauge action \cite{NF2TEST,VALENCE,SEA,SEAVAL,TMQCD}.
 At $\mu=0$ the overall factor for the average plaquette $F_{plaq}$ in
 the parameterization
\be\label{eq3:02}
C_{\tau_{int}} \simeq F\, (am_q)^{-1}\, \Omega
\ee
 turns out to be $F_{plaq} \simeq 2 \cdot 10^7$ in the positive
 quark mass phase and $F_{plaq} \simeq (2-3) \cdot 10^8$ at negative
 quark mass.
 The corresponding numbers at $\mu=0.01$ are between these two values.

 Let us note that at the smallest quark masses a final reweighting
 correction has to be applied because the smallest eigenvalues
 cannot be always kept in the interval of polynomial approximations.
 Sometimes they fluctuate below the lower limit $\epsilon$.

\section{Eigenvalue spectra}\label{sec4}

 Looking at the eigenvalue spectra of the Wilson-Dirac fermion matrix
 (\ref{eq2:01}) at small (untwisted) quark masses (see, for instance, in
 section 4 of ref.~\cite{NF2TEST}) it seems plausible that near zero
 quark mass there has to be a massive rearrangement of eigenvalues.
 This is because in the path integral small eigenvalues are strongly
 supressed by the zero of the fermion determinant.
 At the sign change of the quark mass the eigenvalues have to somehow
 avoid the zero of the determinant at the origin.
 It is plausible that this eigenvalue rearrangement is related to the
 phase transition at zero quark mass.

 An interesting question is how the behaviour of eigenvalues in the
 small quark mass region is influenced by a non-zero twisted mass term.

 We investigated the eigenvalue spectra by the Arnoldi method on
 $8^3\times 16$ and $12^3\times 24$ lattices in some of the runs listed
 in table~\ref{tab_run}.
 Typically 100-200 eigenvalues were determined on 10-30 independent
 gauge field configurations.
 The parameters of the Arnoldi code were set for searching the
 eigenvalues with the smallest absolute value. 

\begin{figure}
\begin{center}
\hspace*{0.05\linewidth}
\begin{minipage}[c]{0.95\linewidth}
 \begin{minipage}[c]{0.65\hsize}
  \includegraphics[angle=-90,width=.9\hsize]
   {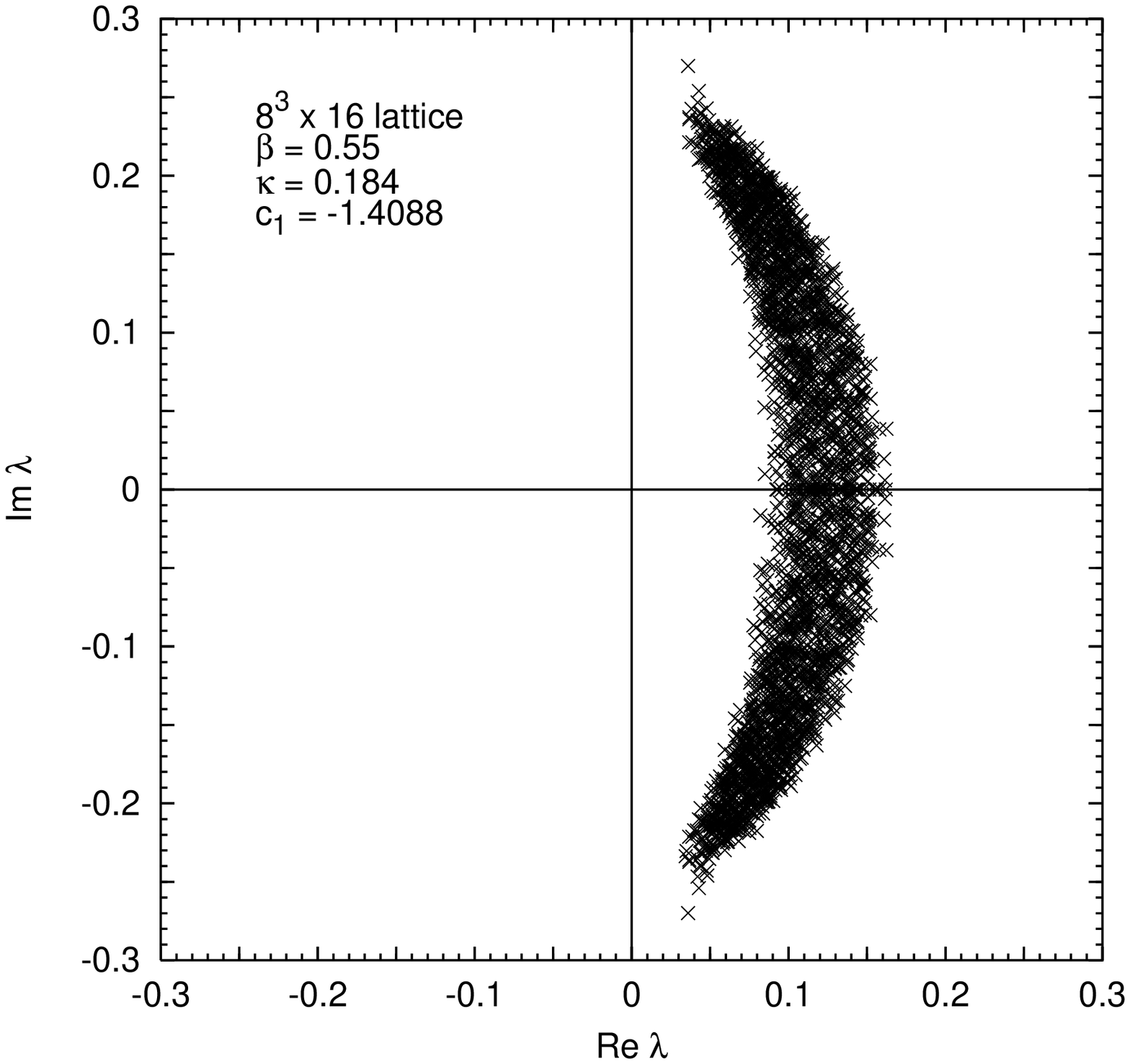}
 \end{minipage}
 \hspace*{-30mm}
 \begin{minipage}[c]{0.65\hsize}
  \includegraphics[angle=-90,width=.9\hsize]
   {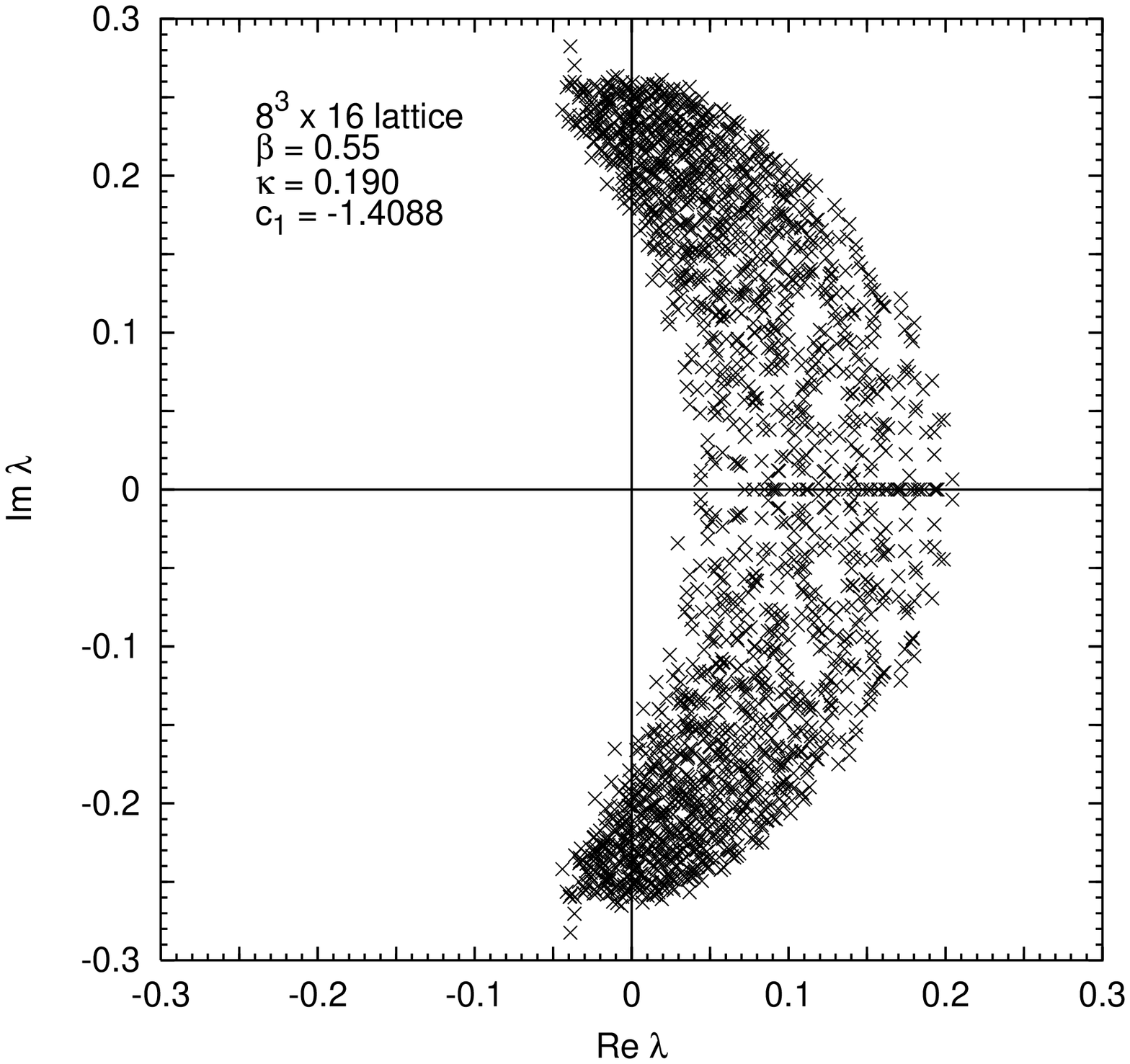}
 \end{minipage}
\end{minipage}
\end{center}
\begin{center}
\hspace*{0.05\linewidth}
\begin{minipage}[c]{0.95\linewidth}
 \begin{minipage}[c]{0.65\hsize}
  \includegraphics[angle=-90,width=.9\hsize]
   {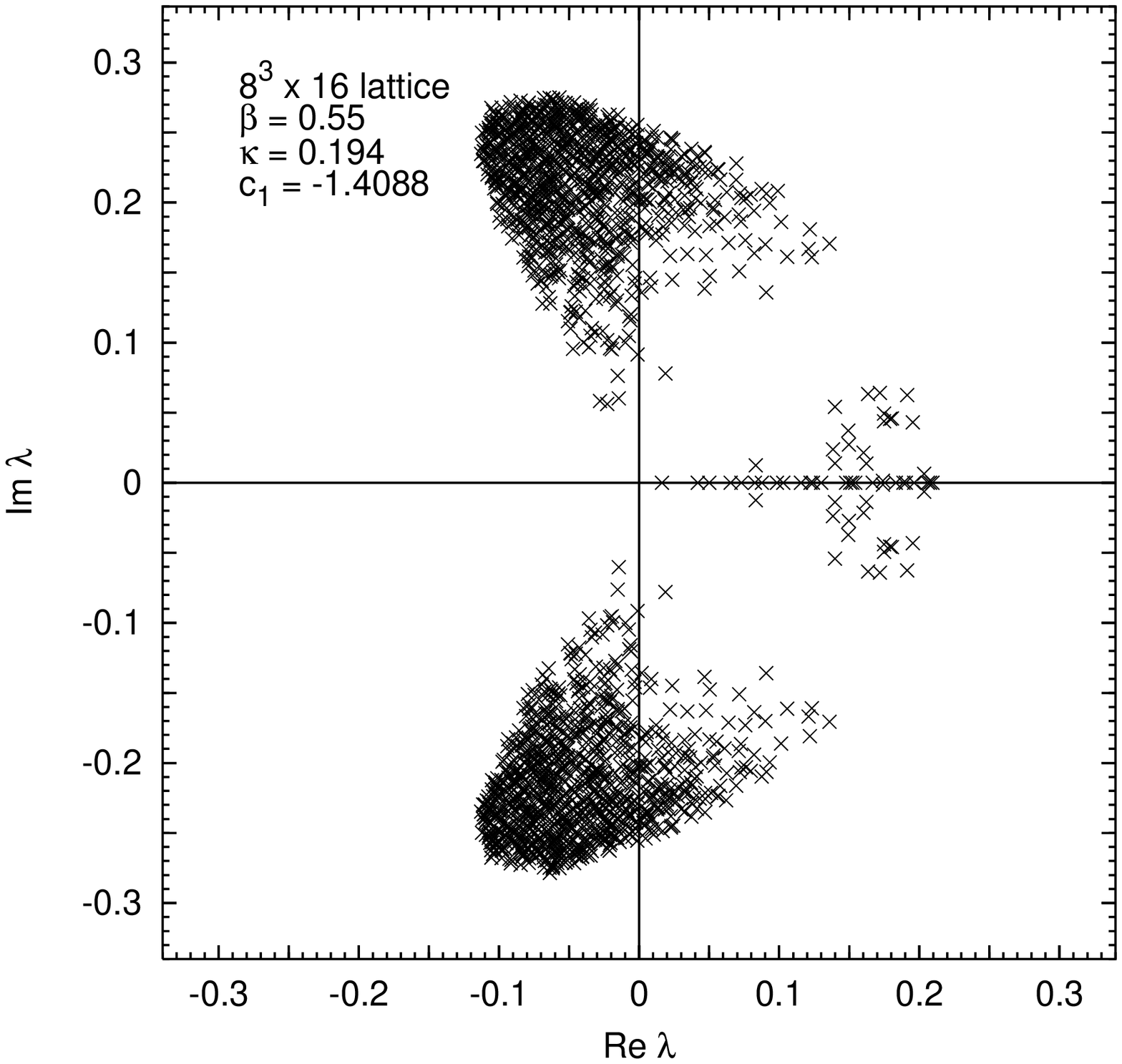}
 \end{minipage}
 \hspace*{-30mm}
 \begin{minipage}[c]{0.65\hsize}
  \includegraphics[angle=-90,width=.9\hsize]
   {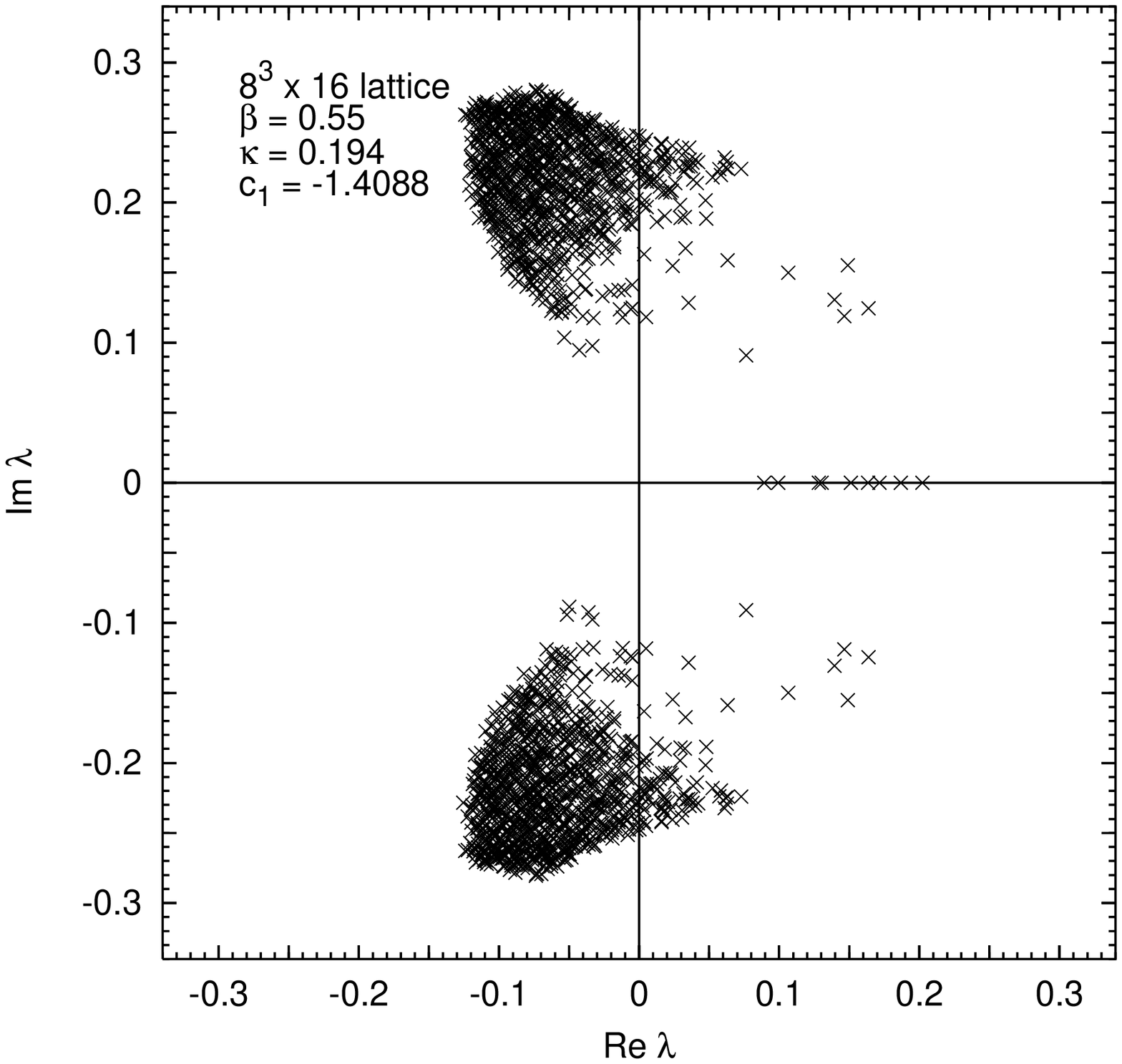}
 \end{minipage}
\end{minipage}
\end{center}
\vspace*{2mm}
\begin{center}
\parbox{0.8\linewidth}{\caption{\label{fig_Eigen8c16}\em
 Eigenvalues of the Wilson-Dirac fermion matrix (\protect\ref{eq2:01})
 with small absolute value for $\mu=0$, $\beta=0.55$ on an
 $8^3\times 16$ lattice.
 Upper left panel: $\kappa=0.184$.
 Upper right pannel: $\kappa=0.190$.
 Lower panels: $\kappa=0.194$ at the beginning of equilibration (left
 panel) and after equilibration (right panel).}}
\end{center}
\end{figure}

 The results at $\mu=0$ on an $8^3\times 16$ lattice are shown by
 figure~\ref{fig_Eigen8c16}.
 In the upper panels of the figure, where the quark mass is positive,
 typical ``half-moons'' filled with eigenvalues can be seen, which
 correspond to the figures in ref.~\cite{NF2TEST}.
 At negative quark mass -- in the lower part of
 figure~\ref{fig_Eigen8c16} -- an almost empty segment in the middle of
 the ``half-moon'' appears.
 Comparing the two figures at negative quark mass one can also see
 how this segment is gradually emptied during equilibration.

 It is remarkable that even after equilibration there are some real
 (``zero-mode``) eigenvalues on the positive axis.
 Our Arnoldi code did not find in these configurations any negative real
 eigenvalues.
 In addition, it is quite surprising that, apart from the empty segment
 in the middle, the half-moon-shaped deformation of the eigenvalue
 region observed at small positive quark masses does not disappear for
 small negative quark masses either.

\begin{figure}
\begin{center}
\hspace*{0.035\linewidth}
 \begin{minipage}[c]
  {1.00\linewidth} 
  \begin{center}
   \includegraphics[angle=-90,width=.65\hsize]
    {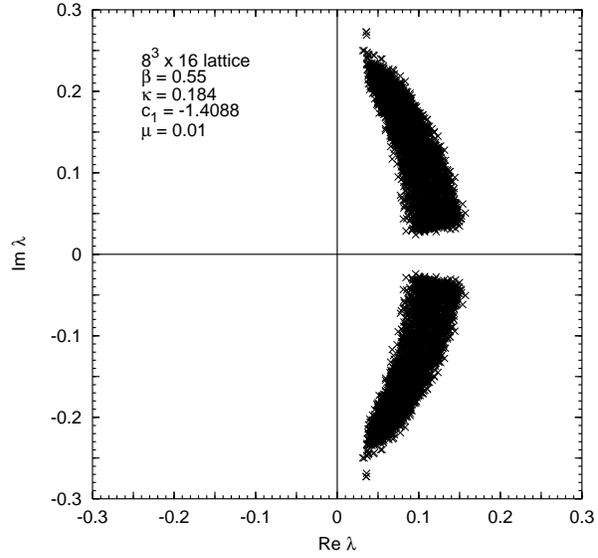}
  \end{center}
  \vspace*{.02\hsize}
  \begin{center}
   \includegraphics[angle=-90,width=.65\hsize]
    {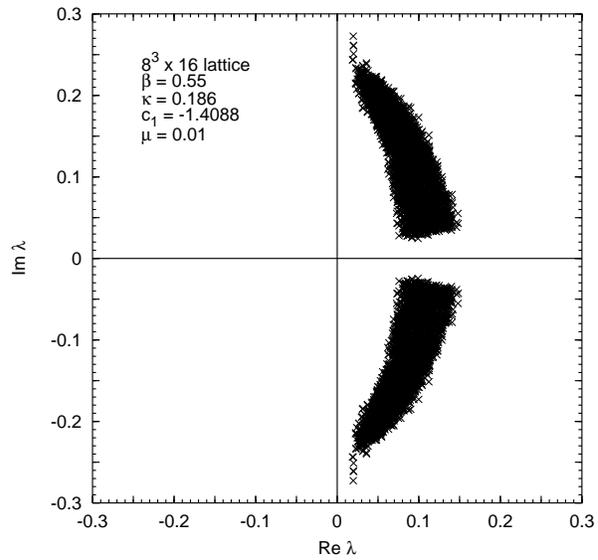}
  \end{center}
 \end{minipage}
\end{center}
\vspace*{-.01\hsize}
\begin{center}
\parbox{0.8\linewidth}{\caption{\label{fig_Eigen8c16_01}\em
 Eigenvalues of the Wilson-Dirac fermion matrix (\protect\ref{eq2:01})
 with small absolute value for $\mu=0.01$, $\beta=0.55$ on an
 $8^3\times 16$ lattice.
 Upper panel: $\kappa=0.184$, lower pannel: $\kappa=0.186$.}}
\end{center}
\end{figure}

 The effect of a non-zero twisted mass on the eigenvalue spectrum
 on an $8^3\times 16$ lattice is illustrated by
 figure~\ref{fig_Eigen8c16_01}.
 It can be seen that the strip around the real axis
 $-\mu \leq {\rm Im}(\lambda) \leq +\mu$ is free from eigenvalues.
 Let us remark that also in presence of a non-zero twisted mass we
 studied the spectrum of the operator of equation (\ref{eq2:01}), which
 corresponds to the so called ``twisted basis''.
 In the ``physical basis'' \cite{FREZZOTTI}, for $\omega=\pi/2$, the
 spectrum of the Dirac operator lies in a vertical line parallel to the
 imaginary axis and is shifted from the origin by $\mu$ (exactly as in
 the continuum).

\begin{figure}
\begin{center}
\hspace*{0.035\linewidth}
 \begin{minipage}[c]
  {1.00\linewidth} 
  \begin{center}
   \includegraphics[angle=-90,width=.65\hsize]
    {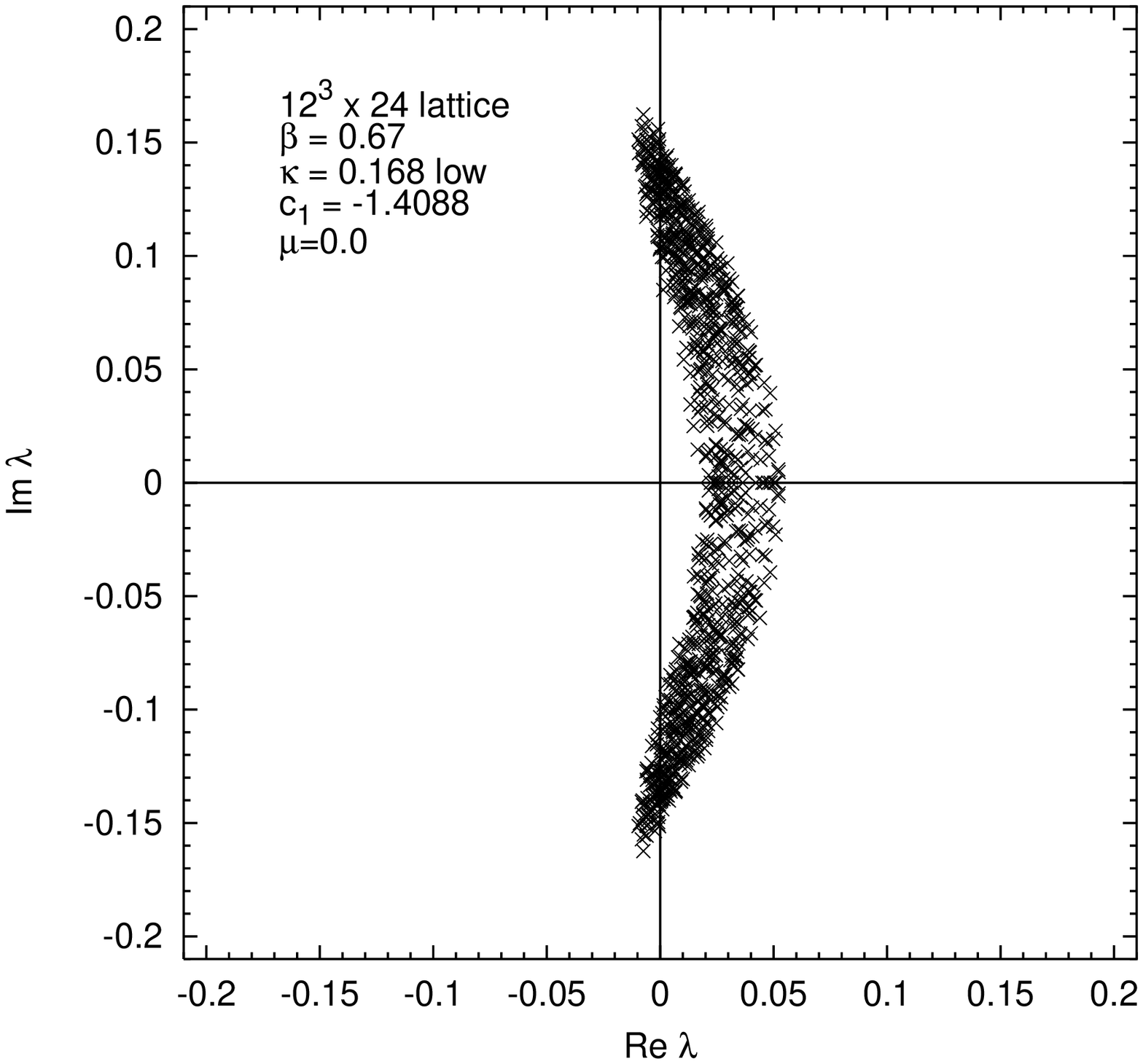}
  \end{center}
  \vspace*{-0.01\hsize}
  \begin{center}
   \includegraphics[angle=-90,width=.65\hsize]
    {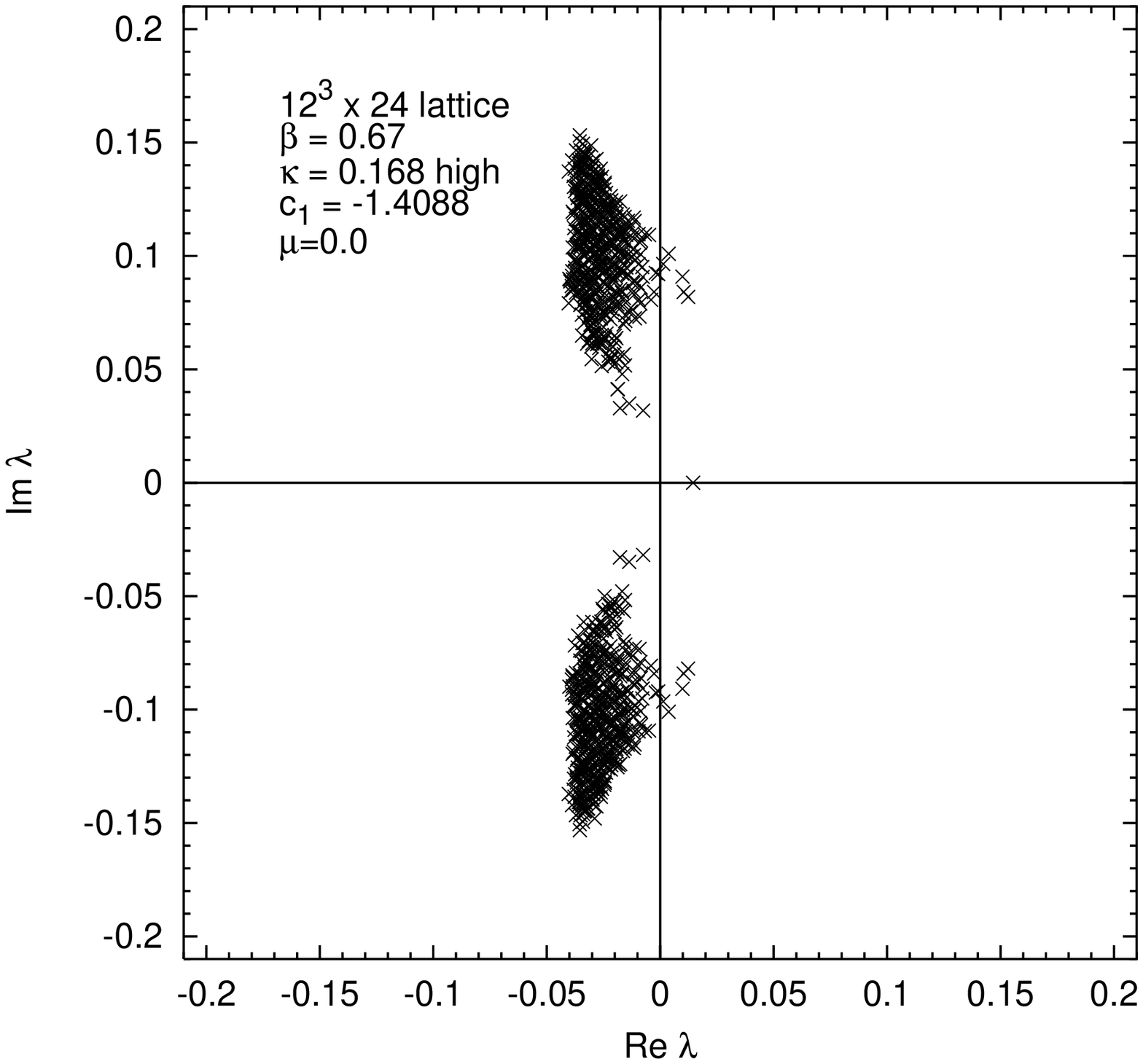}
  \end{center}
 \end{minipage}
\end{center}
\vspace*{-0.01\hsize}
\begin{center}
\parbox{0.8\linewidth}{\caption{\label{fig_Eigen12c24_dbw2}\em
 Eigenvalues of the Wilson-Dirac fermion matrix (\protect\ref{eq2:01})
 with small absolute value at $\beta=0.67,\;\kappa=0.168$ on a
 $12^3\times 24$ lattice.
 Upper panel: $\mu=0$, ``low plaquette'';
 lower pannel: $\mu=0$, ``high plaquette''.}}
\end{center}
\end{figure}

 Going to larger $\beta$ (smaller lattice spacing) the visible
 difference is that the ``half-moons'' are straightened and come
 closer to the origin: see figure~\ref{fig_Eigen12c24_dbw2}.
 Otherwise most qualitative features are unchanged.
 There is, however, a marked difference in the number of real
 eigenvalues (for $\mu=0$): in the upper panels of
 figure~\ref{fig_Eigen8c16} there are lots of them, whereas at
 larger $\beta$, in the upper panel of
 figure~\ref{fig_Eigen12c24_dbw2}, their number is substantially
 reduced.

\begin{figure}
\begin{center}
 \begin{minipage}[c]
  {1.20\linewidth} 
   \begin{center}
   \includegraphics[angle=-90,width=.95\hsize]
    {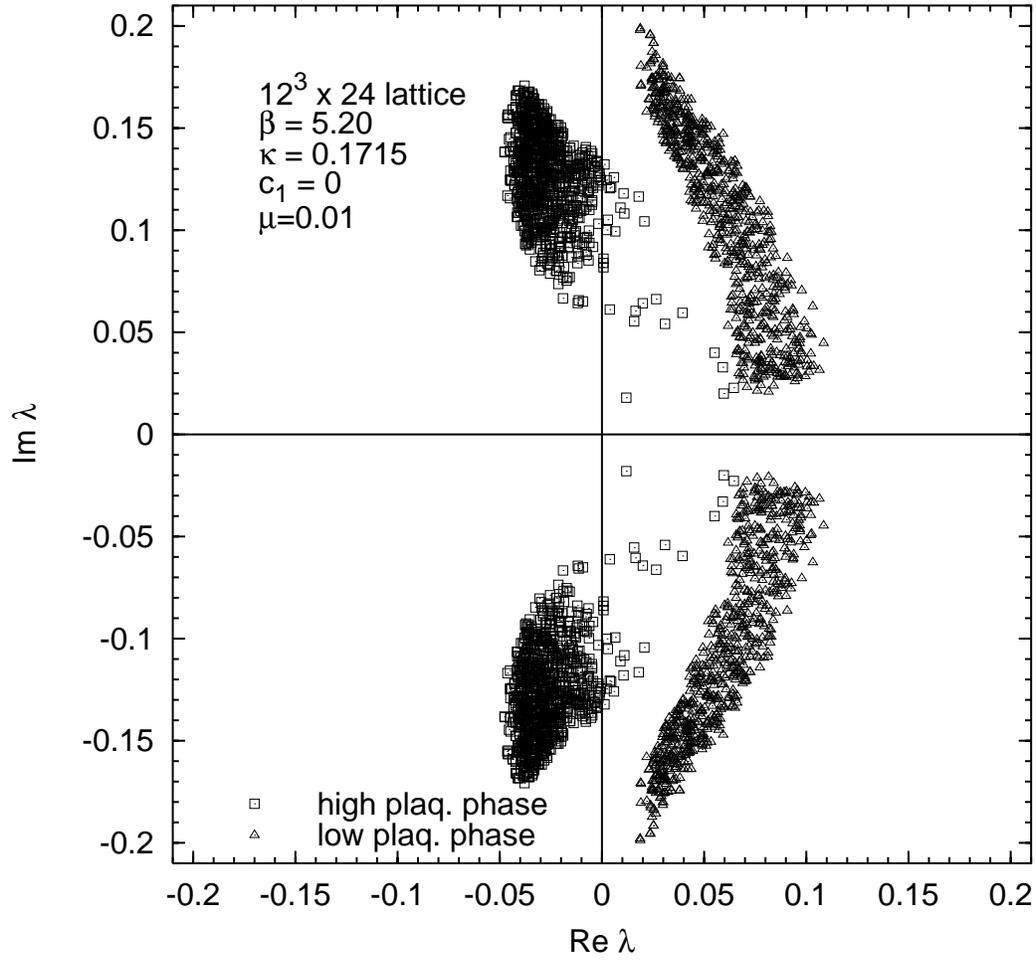}
  \end{center}
 \end{minipage}
\end{center}
\vspace*{-0.01\hsize}
\begin{center}
\parbox{0.8\linewidth}{\caption{\label{fig_Eigen12c24_wilson}\em
 Eigenvalues of the Wilson-Dirac fermion matrix with small absolute
 value in case of the Wilson plaquette action at
 $\beta=5.20,\;\mu=0.01,\;\kappa=0.1715$ on a $12^3\times 24$ lattice.
 Both ``high plaquette'' and ``low plaquette'' spectra are shown.}}
\end{center}
\end{figure}

 The effect of changing the gauge action can be seen by comparing
 figure~\ref{fig_Eigen12c24_dbw2} with the eigenvalue spectra in case
 of the Wilson plaquette gauge action at a similar lattice spacing
 $a \simeq 0.2\,{\rm fm}$ in figure~\ref{fig_Eigen12c24_wilson}.
 The fact that in the case of using the Wilson gauge action the
 pion mass is larger than in the case of the DBW2 action is reflected
 by a movement of the "half-moons" farther away from the origin.

\section{Conclusion}\label{sec5}

 The main conclusion of this paper is that, indeed, exchanging the
 Wilson plaquette gauge action with the (renormalization group improved)
 DBW2 action shows substantial effects on the phase structure:
 We performed a qualitative study of the phase structure of 
 lattice QCD by changing the gauge action and compared the
 Wilson-plaquette and DBW2 actions at a lattice spacing
 $a \simeq 0.2\,{\rm fm}$ in the positive quark mass phase and
 $a \simeq 0.17\,{\rm fm}$ in the phase with negative quark mass.
 This means, at $\beta=5.2$ for the Wilson plaquette action and
 $\beta=0.67$ for the DBW2 action.
 At this comparable situation the metastability signaled by the
 existence of long living states with different average plaquette value
 and quark masses with opposite sign becomes weaker and the minimal pion
 mass and the jump in the average plaquette between the phases with
 positive and negative quark mass decrease.

 For vanishing twisted mass $\mu=0$ the metastability occurs in the
 hopping parameter range $0.167 \leq \kappa \leq 0.168$.
 Going to the twisted mass value $\mu=0.01$, which is the same as
 in the numerical simulations of ref.~\cite{TWIST,FERMILPROC}, the
 metastability disappears on our $12^3\times 24$ lattices.
 It might reappear on larger lattices, but our $12^3\times 24$ data
 indicate that the jump in the average plaquette is at least by a factor
 of ten smaller than the one observed in \cite{TWIST}.

\begin{figure}
\begin{center}
\hspace*{-0.07\vsize}
\begin{minipage}[c]{1.25\linewidth}
\includegraphics[angle=-90,width=.95\hsize]
 {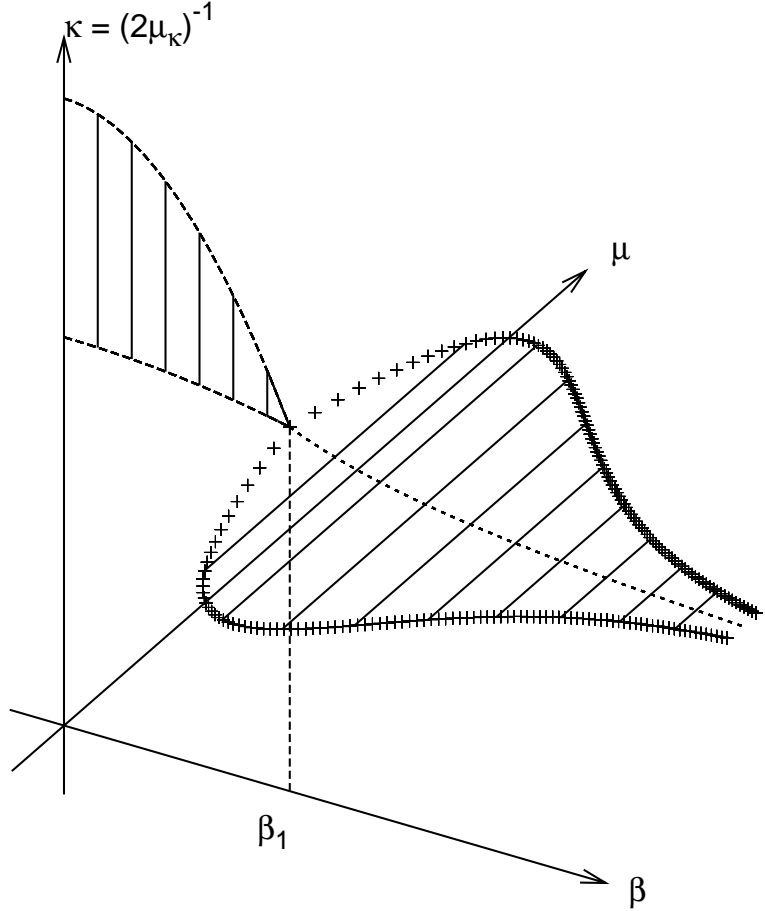}
\end{minipage}
\end{center}
\vspace*{0.01\vsize}
\begin{center}
\parbox{0.8\linewidth}{\caption{\label{fig_phasediagAoki}\em
 The schematic view of the phase transitions in the
 $(\beta,\kappa,\mu)$ space for Wilson quarks with both DBW2 and
 Wilson plaquette gauge action ($\beta$=bare gauge coupling,
 $\kappa$=hopping parameter, $\mu$=bare twisted quark mass,
 $\mu_\kappa \equiv (2\kappa)^{-1}$=bare untwisted quark mass.)
 The crosses mark the second order boundary line of the first order
 phase transition surface.
 At strong gauge coupling there is the surface containing the Aoki
 phase, which ends at a point denoted by $\beta=\beta_1$.
 The figure does not extend down to $\beta=0$ and only one ``finger''
 of the Aoki phase is shown.}}
\end{center}
\end{figure}

 At a lower $\beta$ value $\beta=0.55$, which corresponds to lattice
 spacings $a \simeq 0.30\,{\rm fm}$ and $a \simeq 0.23\,{\rm fm}$
 for positive and negative quark mass, respectively, our simulation
 data are consistent with the existence of the Aoki phase.
 This is similar to the situation for $\beta \leq 4.6$ in case of the
 Wilson plaquette action \cite{HUMBOLDT}.
 The schematic picture of the suggested phase diagram in the
 $(\beta,\kappa,\mu)$ space, both for DBW2 and Wilson plaquette gauge
 actions, is shown by figure~\ref{fig_phasediagAoki}.

 The minimal pion mass in a stable phase can be estimated from our
 simulation data at $\beta=0.67$ and vanishing twisted mass on a 
 $12^3\times 24$ lattice to be $m_\pi^{\rm min} \simeq 250\,{\rm MeV}$
 in the positive quark mass phase and
 $m_\pi^{\rm min} \simeq 375\,{\rm MeV}$ in the phase with negative
 quark mass.
 On larger lattices this value is expected to be 10-20\% smaller due
 to the finite volume effects which are non-negligible on the
 $12^3\times 24$ lattice, especially in the negative quark mass phase
 where the lattice extension is only $L \simeq 2.0\,{\rm fm}$.
 At positive twisted mass $\mu=0.01$ the estimate for the minimal
 charged pion mass is $m_\pi^{\rm min} \simeq 360\,{\rm MeV}$, a value
 entirely due to the non-zero twisted mass and not to the first order
 phase transition.

\begin{figure}
\begin{center}
\hspace*{0.11\linewidth}
\begin{minipage}[c]{1.1\linewidth}
\includegraphics[angle=-90,width=.95\hsize]
 {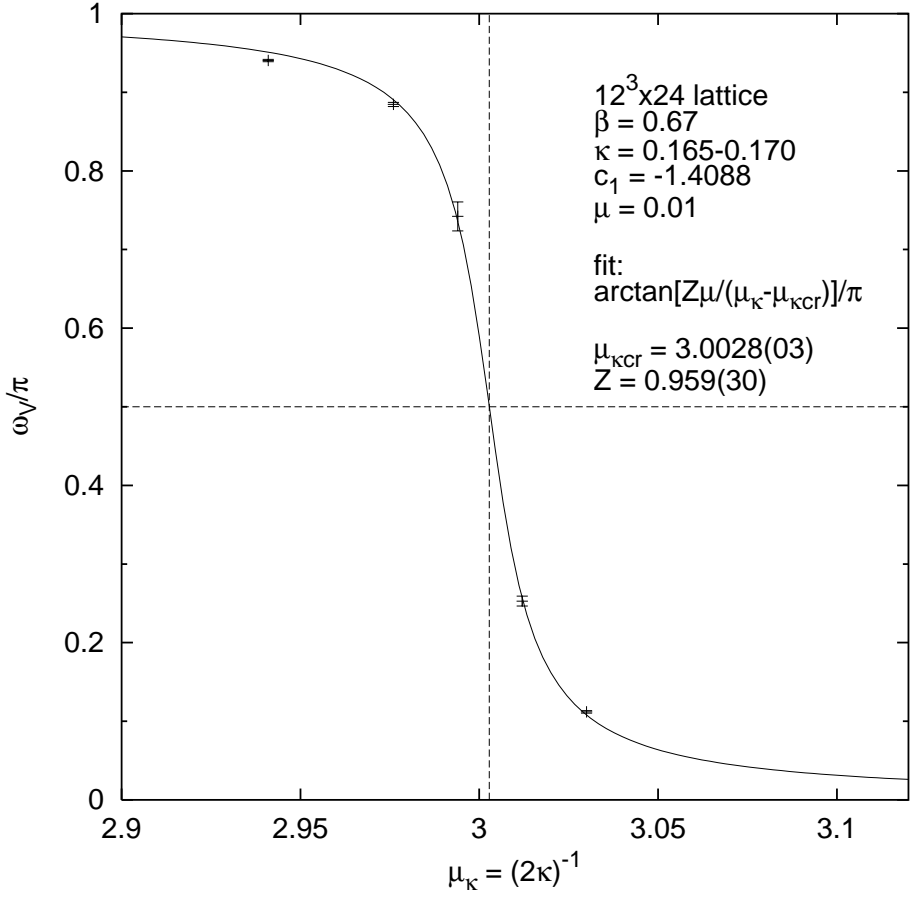}
\end{minipage}
\end{center}
\vspace*{-0.02\vsize}
\begin{center}
\parbox{0.8\linewidth}{\caption{\label{fig_omegav}\em
 The dependence of the twist angle $\omega_V$ on the bare untwisted
 quark mass $\mu_\kappa=(2\kappa)^{-1}$ at $\beta=0.67$, $\mu=0.01$ on a
 $12^3\times 24$ lattice.
 The fit determines the critical hopping parameter to be
 $\kappa_{cr} = (2\mu_{\kappa cr})^{-1} = 0.16651(2)$.}}
\end{center}
\end{figure}

 Besides the pion- and $\rho$-meson masses, at non-vanishing twisted
 mass, we also determined the twist angle $\omega_V$ as a function of
 the bare untwisted quark mass $\mu_\kappa$.
 The $\mu_\kappa$-dependence of $\omega_V$ can be well described by the
 expected $\arctan$-function (see figure~\ref{fig_omegav}).
 From the fit one obtains the value of the critical hopping
 parameter $\kappa_{cr} = 0.16651(2)$ and an estimate of a combination
 of $Z$-factors.

\begin{figure}
\vspace*{-0.04\hsize}
\begin{center}
\hspace*{0.12\linewidth}
\begin{minipage}[c]{1.0\linewidth}
\includegraphics[angle=-90,width=.72\hsize]
 {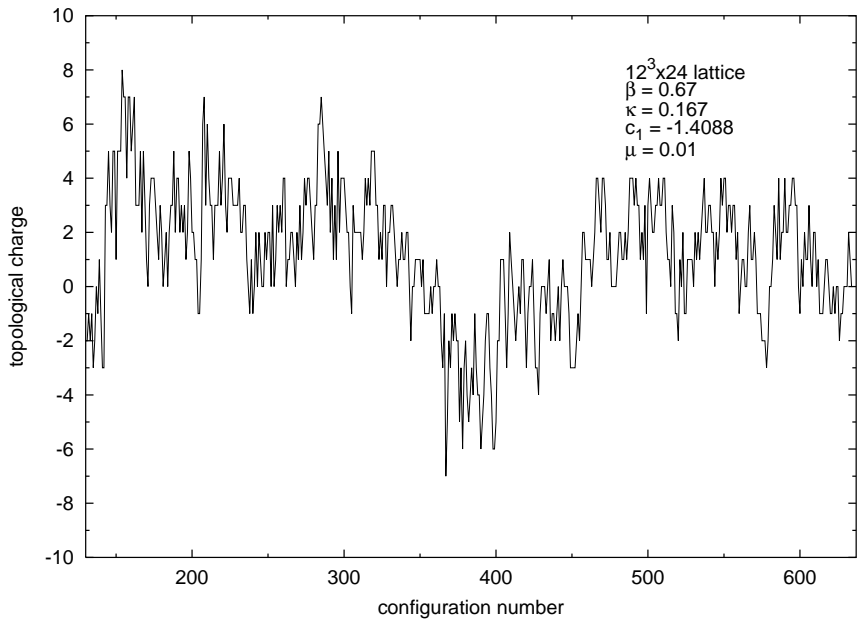}
\end{minipage}
\end{center}
\vspace*{0.03\hsize}
\begin{center}
\hspace*{0.15\linewidth}
\begin{minipage}[c]{1.0\linewidth}
\includegraphics[angle=-90,width=.89\hsize]
 {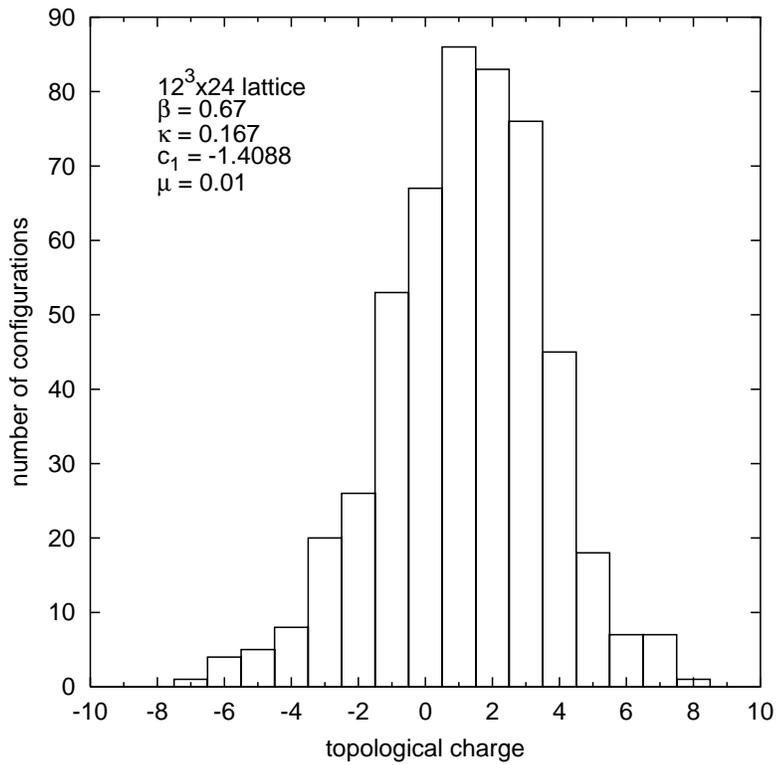}
\end{minipage}
\end{center}
\begin{center}
\parbox{0.8\linewidth}{\caption{\label{fig_Qhist}\em
 The topological charge on a $12^3\times 24$ lattice at $\beta=0.67$,
 $\mu=0.01$, $\kappa=0.167$.
 Upper panel: the run history, with configurations separated by 10 TSMB
 update cycles.
 Lower panel: the distribution of the topological charge in this run.}}
\end{center}
\end{figure}

 In some of the simulation runs we also monitored the history of the
 topological charge (see, for instance, figure~\ref{fig_Qhist}).
 Although the autocorrelation of the topological charge is markedly
 longer than those of the average plaquette or of the pion mass, in
 a good statistics run with, say, thousand times the integrated
 autocorrelation length of the pion mass, the different topological
 sectors could be properly sampled.

 In order to illustrate the rearrangement of the small eigenvalues
 near the zero quark mass phase transition we investigated in some
 detail the eigenvalue spectrum of the non-hermitean fermion matrix
 defined in (\ref{eq2:01}).
 To our surprise, the transition from positive to negative quark
 mass is signaled in the eigenvalue spectrum by the opening up of
 an almost empty segment in the ``half-moon'' occupied by the
 eigenvalues near the origin.
 The introduction of a non-vanishing twisted mass causes the appearance
 of an empty strip $[-\mu,+\mu]$ on both sides of the real axis.
 The effect of larger $\beta$ is to straighten the ``half-moon''
 occupied by the small eigenvalues.
 At the same time the small real eigenvalues at zero twisted mass, which
 are causing the problem of the so-called ``exceptional gauge
 configurations'' in partially quenched simulations, occur much less
 frequently.

 A welcome side-effect of introducing the RGI gauge action is the
 speed-up of the TSMB update algorithm.
 (This presumably also applies to other update algorithms, but in this
 paper we only used TSMB.)
 This can be qualitatively understood by the reduction of the
 probability for small size ``dislocations'' in the gauge field and
 for the less frequent occurrence of small real eigenvalues.
 (This is qualitatively similar to the conclusions of
 ref.~\cite{DHKW}, obtained in another setup.)
 The computational cost as a function of the quark mass can be better
 approximated in the small quark mass region by an inverse power
 behaviour of only $(am_q)^{-1}$ than by the behaviour $(am_q)^{-2}$
 observed previously with the Wilson plaquette action.

 The results of the present paper indicate that the combination of
 $N_f=2$ Wilson-quarks with the DBW2 gauge action leads to a phase
 structure with a weaker first order phase transition than $N_f=2$
 Wilson-quarks with the plaquette gauge action at a comparable value
 of the lattice spacing.
 For the moment we have no detailed information on the dependence
 of the phase structure on the parameter $c_1$ in the gauge action
 which multiplies the rectangular Wilson loops.
 It is possible that the optimal choice is different from
 $c_1=-1.4088$, for instance, $c_1=-0.331$ for to the Iwasaki action
 or $c_1=-1/12$ for the tree-level-improved Symanzik action.
 The best choice of $c_1$ might also be influenced by the positivity
 problem of improved actions \cite{NECCO} and/or by the convergence
 rate of lattice perturbation theory \cite{QCDSF}.

 An important open question, which remains to be investigated in the
 future, is the $\beta$-dependence of the phase structure for
 Wilson-type lattice fermions.
 It is expected that closer to the continuum limit the minimal
 pion mass and the jump in the average plaquette become smaller and
 finally, in the continuum, the first order phase transition line in
 the plane of untwisted and twisted quark mass shrinks to a first order
 phase transition point.
 The faster this actually happens the better it is for
 phenomenologically relevant numerical QCD simulations with Wilson-type
 quarks.

 Another important question is, whether the DBW2 gauge action in
 combination with Wilson twisted mass fermions shows a good scaling
 behaviour.
 To this end, a simulation at a higher value of $\beta$ than the one
 used here is necessary.
 Both questions mentioned above are presently investigated by our
 collaboration.

\vspace*{1em}
\noindent
{\large\bf Acknowledgments}

\noindent
 We thank Roberto Frezzotti, Gernot M\"unster and Giancarlo Rossi for
 helpful discussions, Stanislav Shcheredin for giving us his cooling
 code as well as Silvia Necco and Urs Wenger for their advice and help
 in evaluating the scale parameter $r_0$.
 
 The computations were performed on the IBM-JUMP computer at NIC
 J\"ulich and the PC clusters at DESY Hamburg, NIC Zeuthen,
 Forschungszent\-rum Karlsruhe, University of M\"unster and the Sun Fire
 SMP-Cluster at the Rechenzent\-rum - RWTH Aachen.
 This work was supported by the DFG Sonderforschungsbereich/Transregio
 SFB/TR9-03.


\newpage

\begin{table}
\begin{center}
{\Large\bf Tables}
\end{center}
\vspace*{1em}
\begin{center}
\parbox{0.8\linewidth}{\caption{\label{tab_run}\em
 Bare couplings and parameters of the TSMB algorithm in runs with the
 DBW2 gauge action.
 The determinant breakup multiplicity is $n_B=4$ in all runs.
 Small letters label runs on $8^3\times 16$ lattices at $\beta=0.55$
 whereas capital letters stand for runs on $12^3\times 24$ lattices
 at $\beta=0.67$.
 The suffix $l$ and $h$ denote ``low'' and ``high'' plaquette phase,
 respectively. 
 Those runs with a caligraphic letter are performed with an additional
 twisted mass term ($\mu=0.01$).
 The number of analyzed configurations is given in the last column.
 An asterix on these numbers denotes that a few configurations have very
 low ($\ll 1$) reweighting factors.
 The analyzed gauge configurations are separated by 10 update cycles,
 except for run $(a)$ where they are separated by 100 update cycles.}}
\end{center}
\begin{center}
\begin{tabular}{|c|c|c|c|c|c|c|r|}
\hline
run & $\kappa$  & $n_1$ & $n_2$ & $n_3$ & $\lambda$ & $\epsilon$ &
$N_{conf}$
\\ \hline
$(a)$ & 0.184 & 22 & 100 & 102 & 24 & 2.4$\cdot 10^{-3}$ & 116
\\ \hline
$(b)$ & 0.186 & 22 & 200 & 220 & 23 & 5.8$\cdot 10^{-4}$ & 381
\\ \hline
$(c)$ & 0.188 & 24   & 500 & 520 & 23 & 5.7$\cdot 10^{-5}$ & 165 
\\ \hline
$(d)$ & 0.190 & 30 & 900 & 940 & 22 & 1.1$\cdot 10^{-5}$ & 66$^\ast$
\\ \hline
$(e)$ & 0.192 & 30 & 1400 & 1440 & 22 & 2.7$\cdot 10^{-6}$ & 159$^\ast$
\\ \hline
$(f)$ & 0.193 & 26 & 650 & 680 & 22 & 2.7$\cdot 10^{-5}$ & 192
\\ \hline
$(g)$ & 0.194 & 22 & 300 & 320 & 21 & 2.1$\cdot 10^{-4}$ & 111
\\ \hline\hline
$(A_{\mbox{l}})$ & 0.165 & 28 & 210 & 220 & 26  & $1.3\cdot 10^{-3}$ & 82 
\\ \hline
$(C_{\mbox{l}})$ & 0.167 & 28 & 500 & 510 & 25  & $1.3\cdot 10^{-4}$ & 62
\\ \hline
$(C_{\mbox{h}})$ & 0.167 & 30 & 1100 & 1200 & 25 & $1.3\cdot 10^{-5}$ & 220
\\ \hline
$(D_{\mbox{l}})$ & 0.168 & 30 & 1100 & 1200 & 25 & $1.2\cdot 10^{-5}$ & 82$^\ast$
\\ \hline
$(D_{\mbox{h}})$ & 0.168 & 30 & 1100 & 1200 & 25 & $1.2\cdot 10^{-5}$ & 211
\\ \hline
$(E_{\mbox{h}})$ & 0.170 & 28 & 900 & 920 & 24 & $4.8\cdot 10^{-5}$ & 194
\\ \hline
$(F_{\mbox{h}})$ & 0.172 & 28 & 500 & 510 & 24  & $1.2\cdot 10^{-4}$ & 151
\\ \hline
$(G_{\mbox{h}})$ & 0.175 & 28 & 500 & 510 & 23  & $1.1\cdot 10^{-4}$ & 78$^\ast$
\\ \hline\hline
$(\mathcal{A}_{\mbox{l}})$ & 0.165 & 16 & 250 & 270 & 24 & $1.2\cdot 10^{-3}$  & 540
\\ \hline
$(\mathcal{B}_{\mbox{l}})$ & 0.166  & 18 & 420 & 460 & 24 & $3.6\cdot 10^{-4}$ & 58
\\ \hline
$(\mathcal{C}_{\mbox{h}})$ & 0.167  & 18 & 420 & 460 & 24 & $3.6\cdot 10^{-4}$ & 139
\\ \hline    
$(\mathcal{D}_{\mbox{h}})$ & 0.168  & 18 & 420 & 460 & 24 & $3.6\cdot 10^{-4}$ & 321
\\ \hline    
$(\mathcal{E}_{\mbox{h}})$ & 0.170  & 18 & 420 & 460 & 24 & $3.6\cdot 10^{-4}$ & 100
\\ \hline   
\end{tabular}
\end{center}
\end{table}

\begin{table}
\begin{center}
\parbox{0.8\linewidth}{\caption{\label{tab_resu}\em
 Results of runs specified in table \protect{\ref{tab_run}} for
 different quantities.}}
\end{center}
\begin{center}
\begin{tabular}{|c|l|l|l|l|l|l|l|}
\hline
\multicolumn{1}{|c|}{$\mbox{run}$}      &
\multicolumn{1}{|c|}{$am_\pi$}          &
\multicolumn{1}{|c|}{$am_\rho$}         &
\multicolumn{1}{|c|}{$m_\pi/m_\rho$}    &
\multicolumn{1}{|c|}{$am_\chi^{PCAC}$}
\\ \hline
$(a)$ & 0.6962(69) & 1.0015(75) & 0.6952(37) 
 & 0.07086(85)
\\ \hline
$(b)$ & 0.5325(60) & 0.9013(75) & 0.5908(57) 
 & 0.03890(75)
\\ \hline
$(c)$ & 0.3652(49) & 0.840(26) & 0.435(13) 
 & 0.0154(10)
\\ \hline
$(d)$ & 0.081(24) & 0.62(38) & 0.130(78) 
 & 0.0012(15)
\\ \hline
$(e)$ & 0.594(51) & 1.80(30) & 0.355(42) 
 & -0.0430(70)
\\ \hline
$(f)$ & 0.888(19) & 1.794(30) & 0.495(13) 
 & -0.0870(38)
\\ \hline
$(g)$ & 0.997(23) & 1.820(59) & 0.548(17) 
 & -0.0995(66)
\\ \hline\hline
$(A_{\mbox{l}})$ & 0.454(04) & 0.724(25) & 0.627(18) 
 & 0.0414(05)
\\ \hline
$(C_{\mbox{l}})$  & 0.343(07) & 0.735(32) & 0.466(21) 
 & 0.0222(11)
\\ \hline
$(C_{\mbox{h}})$  & 0.313(22) & 0.776(125) & 0.403(67) 
 & -0.0222(28)
\\ \hline
$(D_{\mbox{l}})$  & 0.153(12) & 0.445(109) & 0.344(91) 
 & 0.0053(17)
\\ \hline
$(D_{\mbox{h}})$  & 0.380(31) & 1.144(88) & 0.332(37) 
 & -0.0335(54)
\\ \hline
$(E_{\mbox{h}})$  & 0.644(15) & 1.324(75) & 0.487(27)  
 & -0.0834(38)
\\ \hline
$(F_{\mbox{h}})$  & 0.840(23) & 1.468(52) & 0.572(25) 
 & -0.1295(77)
\\ \hline
$(G_{\mbox{h}})$  & 1.005(44) & 1.801(81) & 0.558(28) 
 & -0.1585(103)
\\ \hline\hline
$(\mathcal{A}_{\mbox{l}})$ & 0.4641(45) & 0.7228(58) & 0.6421(53)  
 & 0.03803(81)
\\ \hline
$(\mathcal{B}_{\mbox{l}})$ & 0.341(05) & 0.634(55) & 0.538(45)  
 & 0.0177(22)
\\ \hline
$(\mathcal{C}_{\mbox{h}})$ & 0.291(12) & 0.607(232) & 0.480(178) 
 & -0.0149(22)
\\ \hline
$(\mathcal{D}_{\mbox{h}})$ & 0.472(07) & 1.035(72) & 0.456(32) 
 & -0.0469(16)
\\ \hline
$(\mathcal{E}_{\mbox{h}})$ & 0.712(14) & 1.136(65) & 0.627(34) 
 & -0.0946(72)
\\ \hline    
\end{tabular}
\end{center}
\end{table}

\begin{table}
\begin{center}
\parbox{0.8\linewidth}{\caption{\label{tab_auto}\em
 The cost of an update cycle $C_{cycle}$ in thousands of MVMs according
 to eq.~(\protect\ref{eq2:17}) and the estimated integrated
 autocorrelation lengths in update cycles obtained from runs specified
 by table \protect{\ref{tab_run}}.
 The suffix $plaq$ and $m_\pi$ refer to the average plaquette and the
 pion mass, respectively.
 The last two columns give the factors $F$ calculated from
 eq.~(\protect\ref{eq2:19}) with $z=2$.}}
\end{center}
\begin{center}
\begin{tabular}{|c|c|r|r|c|c|}
\hline
\multicolumn{1}{|c|}{$\mbox{run}$}         &
\multicolumn{1}{|c|}{$C_{cycle}$}          &
\multicolumn{1}{|c|}{$\tau_{int}^{plaq}$}  &
\multicolumn{1}{|c|}{$\tau_{int}^{m_\pi}$} &
\multicolumn{1}{|c|}{$F_{plaq}/10^6$}      &
\multicolumn{1}{|c|}{$F_{m_\pi}/10^6$}
\\ \hline
$(a)$ & 13 & 152 &  &  11.9 & 
\\ \hline
$(b)$ & 19 & 100 & 20 &  3.5 & 0.7
\\ \hline
$(c)$ & 30 & 147 & $<$ 5 &  1.3 & $<$ 0.04
\\ \hline
$(d)$ & 48 &  & 12 &   & 0.001 
\\ \hline
$(e)$ & 65  & 167 & $<$ 5 & 24  & $<$ 0.7
\\ \hline
$(f)$ & 38 & 95 & 9 & 33  & 3.1
\\ \hline
$(g)$ & 25 & 32 & $<$ 5 & 9.5  & $<$ 1.5
\\ \hline\hline
$(A_{\mbox{l}})$ & 19 & 21  & $<$ 5 & 0.8  & $<$ 0.2
\\ \hline
$(C_{\mbox{l}})$ & 29 & 18  & 15    & 0.3  & 0.3
\\ \hline
$(C_{\mbox{h}})$ & 50 & 53  & 33    & 1.5  & 0.9
\\ \hline
$(D_{\mbox{l}})$ & 51 & 77  & $<$ 5 & 0.1  & $<$ 0.01
\\ \hline
$(D_{\mbox{h}})$ & 51 & 113 & 7     & 7.8  &  0.5
\\ \hline
$(E_{\mbox{h}})$ & 43 & 61  & 11    & 22   &  3.9
\\ \hline
$(F_{\mbox{h}})$ & 30 & 56  & $<$ 5 & 33.4 & $<$ 3.0
\\ \hline
$(G_{\mbox{h}})$ & 31 & 52  & 6     & 48.4 & 5.6
\\ \hline\hline
$(\mathcal{A}_{\mbox{l}})$  & 12 & 143 & 13 & 5.9  & 0.5
\\ \hline
$(\mathcal{B}_{\mbox{l}})$  & 21 & 41  & 9  & 0.6  & 0.1 
\\ \hline    
$(\mathcal{C}_{\mbox{h}})$  & 21 & 22  & 6  & 0.2  & 0.1
\\ \hline
$(\mathcal{D}_{\mbox{h}})$  & 21 & 72  & 8  & 7.8  & 0.9
\\ \hline
$(\mathcal{E}_{\mbox{h}})$  & 21 & 29  & 7  & 12.8 & 3.1
\\ \hline
\end{tabular}
\end{center}
\end{table}

\newpage

\end{document}